\documentclass[preprintnumbers,superscriptaddress,nofootinbib,aps,prd,floatfix]{revtex4}

\usepackage{hyperref}
\usepackage{amsmath,slashed}
\usepackage{graphicx}

\newcommand{\be}{\begin{eqnarray*}}
\newcommand{\ee}{\end{eqnarray*}}

\newcommand{\bee}{\begin{eqnarray}}
\newcommand{\eee}{\end{eqnarray}}
\newcommand{\beeq}{\begin{equation}}
\newcommand{\eeeq}{\end{equation}}

\newcommand{\GeV}{~GeV}

\def\spa#1.#2{\left\langle#1#2\right\rangle}
\def\spb#1.#2{\left[#1#2\right]}
\def\lor#1.#2{\left(#1#2\right)}
\def\sand#1.#2.#3{%
\left\langle\smash{#1}{\vphantom1}^{-}\right|{#2}%
\left|\smash{#3}{\vphantom1}^{-}\right\rangle}
\def\sandp#1.#2.#3{%
\left\langle\smash{#1}{\vphantom1}^{-}\right|{#2}%
\left|\smash{#3}{\vphantom1}^{+}\right\rangle}
\def\sandpp#1.#2.#3{%
\left\langle\smash{#1}{\vphantom1}^{+}\right|{#2}%
\left|\smash{#3}{\vphantom1}^{+}\right\rangle}
\def\sandpm#1.#2.#3{%
\left\langle\smash{#1}{\vphantom1}^{+}\right|{#2}%
\left|\smash{#3}{\vphantom1}^{-}\right\rangle}
\def\sandmp#1.#2.#3{%
\left\langle\smash{#1}{\vphantom1}^{-}\right|{#2}%
\left|\smash{#3}{\vphantom1}^{+}\right\rangle}
\def\spab#1.#2.#3{\langle#1|#2|#3]}
\def\spba#1.#2.#3{[#1|#2|#3\rangle}
\def\spaa#1.#2.#3{\langle#1|#2|#3\rangle}
\def\spbb#1.#2.#3{[#1|#2|#3]}
\def\spaxa#1.#2.#3.#4{\langle#1|#2|#3|#4\rangle}
\def\spbxb#1.#2.#3.#4{[#1|#2|#3|#4]}

\DeclareGraphicsExtensions{.pdf,.png,.jpg}

\usepackage{color}

\begin{document} 

\title{Constraining Dark Sectors at Colliders: Beyond the Effective Theory Approach}

\begin{abstract}
\noindent We outline and investigate a set of benchmark simplified models with the aim of providing a minimal 
simple framework for an interpretation of the existing and forthcoming searches of dark matter particles at the LHC. 
The simplified models we consider provide microscopic QFT descriptions of interactions between the Standard Model
partons and the dark sector particles mediated by the four basic types of messenger fields: scalar, 
pseudo-scalar, vector or axial-vector. Our benchmark models are characterised by four to five parameters,
including the mediator mass and width, the dark matter mass and an effective coupling(s). In the gluon fusion production channel 
we resolve the top-quark in the loop and compute full top-mass effects for scalar and pseudo-scalar messengers.
We show the LHC limits and reach at 8 and 14 TeV for models with all four messenger types. 
We also outline the complementarity of direct detection, indirect detection and LHC bounds for dark matter searches.
Finally, we investigate the effects which arise from extending the simplified model to include potential 
new physics contributions in production. Using the scalar mediator as an example we study the impact of heavy new physics 
loops which interfere with the top mediated loops. 
Our computations are performed within the MCFM framework and we provide fully flexible public Monte Carlo implementation.
\end{abstract}

\author{Philip Harris}

\affiliation{CERN, CH-1211 Geneva 23, Switzerland }
\author{Valentin V. Khoze}
\affiliation{Institute for Particle Physics Phenomenology, Department
  of Physics,\\Durham University, Durham DH1 3LE, United Kingdom}
\author{Michael Spannowsky}
\affiliation{Institute for Particle Physics Phenomenology, Department
  of Physics,\\Durham University, Durham DH1 3LE, United Kingdom}
\author{Ciaran Williams}
\affiliation{Niels Bohr Institute, University of Copenhagen, Blegdamsvej 17, DK-2100 Copenhagen, Denmark}
\affiliation{Department of Physics, University at Buffalo \\
The State University of New York, 
Buffalo, NY 14260-1500, USA\\
\smallskip
{\tt philip.coleman.harris@cern.ch, valya.khoze@durham.ac.uk, michael.spannowsky@durham.ac.uk, ciaran@nbi.dk}}

\preprint{IPPP/14/94}
\preprint{DCPT/14/188}

\maketitle

\section{Introduction}
\label{sec:intro}

Many extensions of the Standard Model predict the existence of dark particles that can be produced in collisions of ordinary matter, but are not directly measurable at the LHC's multipurpose experiments. When measuring the radiation in a scattering event, dark particles would manifest themselves as missing transverse energy i.e. as an imbalance of the total transverse momentum in the event. Dark particles constitute the dark sector which plays a special role in any comprehensive beyond the Standard Model (BSM) description of particle physics: some of the 
dark sector particles can be cosmologically stable\footnote{This is usually a result of of an automatic or imposed discrete symmetry 
present in the dark sector.} and as such they give rise to the dark matter. Non-trivial dark sectors can also contain massless
vector fields which contribute to the dark radiation in the Universe. The fact that the dark matter (DM) and dark radiation 
give the dominant contributions to the matter and radiation in the Universe is one of the clearest indications that 
the Standard Model is incomplete and new physics effects must be present 
in a more fundamental theory of particle interactions.

While dark sectors can be complex, resulting in a rich and varied BSM phenomenology, they have one universal feature which is of particular importance: dark matter particles can interact with visible matter by exchanging a mediator field. When studying scenarios for the production of dark particles at colliders, we consider processes in which a mediator
is produced initially in the course of the hadron-hadron collision. This mediator then subsequently decays, either back to SM degrees of freedom or into the dark sector particles. These latter channels will correspond to events with missing transverse energy at colliders. There is no a priori
requirement that the mediator decays directly into cosmologically stable DM; all decays into dark particles which are stable at collider scales or 
do not result in measured displaced vertices will manifest themselves as missing energy. In this sense it is the production 
and the role of the mediator particle(s) which is of key importance in the collider searches for dark sectors; the actual dark matter 
is a derivative. 

Depending on the nature of the mediator field (arguably the most interesting choices being a vector, axial-vector, scalar or a pseudo-scalar) different mediator production mechanisms can occur. For vectors and axial-vectors, 
a common assumption is that the dominant production mechanism is the
quark-antiquark annihilation at tree-level.  For scalars and pseudo-scalars on the other hand, 
the gluon fusion processes are more relevant. This is inspired by the recent Higgs discovery and assumes that the coupling strength of the new scalars to Standard Model fermions is proportional to their SM Yukawa couplings. 

\begin{center}
\begin{figure}[h,t]
\includegraphics[width=0.25\textwidth]{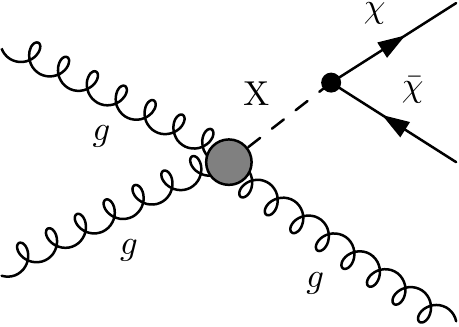} \hspace{0.5cm}
\includegraphics[width=0.3\textwidth]{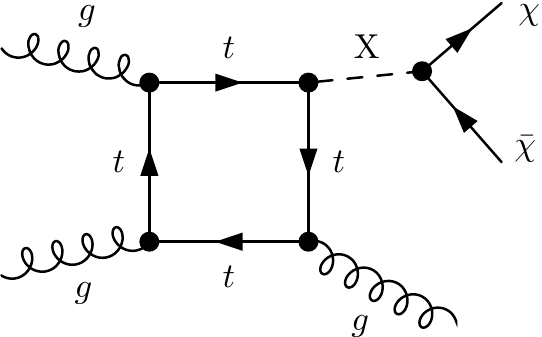} \hspace{0.5cm} 
\includegraphics[width=0.28\textwidth]{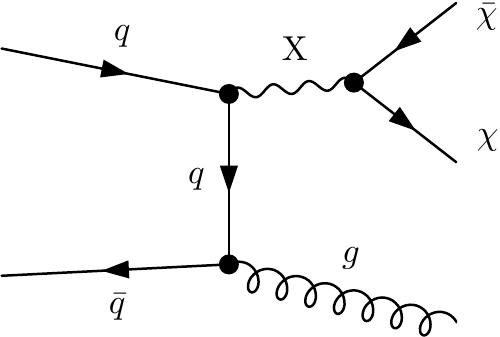}
\caption{Representative Feynman graphs for gluon and quark induced mono-jet processes. The particle X can be either a scalar, pseudo-scalar, vector or axial-vector mediator. The left diagram shows an effective operator approximation of the mediator coupling to gluons. The middle graph represents the full description of the same process, including the fermion mass dependence in the loop, while the right graph depicts a mediator produced in a quark-anti-quark annihilation.
}
\label{fig:feyn}
\end{figure}
\end{center}

Inferring the existence of dark particles in collider experiments requires them to recoil against visible radiation. 
Since the recoil object need not be essential in the interaction which produces the mediator, a natural candidate for the tagging object is the emission of initial state radiation, which occurs at a high rate at the LHC.
In these mono-jet signatures a hard jet recoils against the invisible particles. Events with several hard jets are often vetoed, leaving $Z/W$+jets as major Standard Model backgrounds. In these events the transverse momentum of the jet sets the energy scale of the hard interaction.  

The mediating particle can couple directly or indirectly to the initial state patrons, representative diagrams 
are shown in Fig.~\ref{fig:feyn}. The diagram on the far right of Fig.~\ref{fig:feyn} represents an example in which the mediator-SM 
interaction proceeds via a tree-level interaction with quarks. The mediator can also couple to initial states indirectly, in these instances 
the underlying production mechanism corresponds to loop-induced process, the middle diagram in Fig.~\ref{fig:feyn} illustrates this scenario. 
The propagating loop particle can be integrated out, resulting in an effective dimension-5 operator, illustrated in the left most diagram of Fig.~\ref{fig:feyn}. This prescription is invalid if $p_{T,j} \gtrsim \mathcal{O}(m_X)$, where $m_X$ is the mass of the loop particle. In the case of the top quark, this can readily be achieved.  On the other hand, heavy colored states which couple to the mediator can be integrated out provided  $\Lambda_{\mathrm{NP}}$ is much larger than the energy scale where the operator is probed.


To be able to probe new physics models with particle masses below the characteristic interaction scale of the hard interaction, so-called simplified models were proposed \cite{Alves:2011wf} which only make assumptions on the quantum numbers of particles involved in the minimal processes at the microscopic level, thereby correctly capturing the kinematic features of the new physics model. 

\medskip

The simplified model framework for dark matter and dark sector searches at colliders should constitute a list of key relevant QFT interactions which first produce a mediator particle in a proton-proton collision which subsequently decays into 
other particles, including dark matter. In general such benchmark models would be characterised by the production mechanism (e.g.
$q \bar{q}$ or gluon gluon, etc), the type of the mediator (e.g. scalar, pseudo-scalar, vector or axial-vector) and the 
decay channel (e.g. $s$-channel or $t$-channel production of two dark matter fermions, or other DM particle species).
Secondly, each individual class of these simple models should be characterised by an appropriately chosen minimal set of physically relevant parameters
(coupling constants, masses and widths).

The uses of the simplified model approach in the context of mono-jets and mono-photons searches at colliders and the discussion of its scope have become particularly relevant now in the light of the forthcoming 
run 2 of the LHC. The emerging framework is attracting a fair amount of attention in the collider and phenomenology communities. 
Two recent overviews \cite{Abdallah:2014hon,Malik:2014ggr} give an example of this. 
The aim of the present paper is to go beyond the Born-level processes of dark matter production in the quark-anti-quark channel
and include processes with gluons in the initial state. 

The authors of Ref.~\cite{Abdallah:2014hon} have discussed examples of tree-level benchmark processes relevant for 
interpreting DM searches
at colliders, specifically:
quark-anti-quark $s$-channel processes mediated by Scalar (S)$^s_{q\bar{q}}$ and Vector (V)$^s_{q\bar{q}}$ messengers; 
and the $t$-channel processes mediated by Colored Scalar (CS)$^t_{q\bar{q}}$ messengers. They have also considered
gluon fusion via dimension-5 EFT operators mediated by Scalar (S)$^{\rm EFT}_{gg}$
and Pseudo-scalar messengers (P)$^{\rm EFT}_{gg}$ and have commented on EFT models in which DM coupled preferentially to the third generation.

We will extend these considerations by computing gluon fusion processes at 1-loop-level in a microscopic theory and apply this analysis to simplified DM models with scalar (S) and pseudo-scalar (P) mediators. 
 To enable a direct comparison between models with different mediator types for the LHC reach,
 we will also re-evaluate the predictions of vector (V) and axial-vector (A) mediators produced in the quark-anti-quark channel.
 
 Our work is also complimentary to the recent White Paper \cite{Malik:2014ggr} which has concentrated specifically on the
 cases of vector and axial-vector messengers in the ${q\bar{q}}$ channel. More generally,
 i.e. going beyond the point of which specific mediator-types are included, Ref.~\cite{Malik:2014ggr}  
 has addressed an important task of identifying the minimal number of relevant parameters characterising simplified models for DM collider searches. Their proposal is to select four parameters: the DM mass $m_{\rm DM}$,
 the mediator mass $m_{\rm MED}$, and the two coupling constants: $g_{\rm SM}$ characterising the coupling of the SM quarks 
 to the mediator, and $g_{\rm DM}$ which is the coupling of the mediator to dark particles.
 
 We would like to emphasise here the importance of the mediator width $\Gamma_{\rm MED}$ which we will treat as
 an independent parameter inherent in the characterisation of the simplified models. The impact of $\Gamma_{\rm MED}$  
 in mono-jet searches has been already discussed in earlier literature, see e.g. \cite{Fox:2011fx, Fox:2011pm,Fox:2012ru,An:2012va,Buchmueller:2013dya}.
 The DM production cross section at colliders scales numerically as 
 $\sigma \propto g_{\rm SM}^2 g_{\rm DM}^2 /(m_{\rm MED}^4 \Gamma_{\rm MED}),$ i.e. inversely proportional to the width, leading to 
 a resonant enhancement of the cross section at small values of $\Gamma_{\rm MED}$ as pointed out in \cite{Malik:2014ggr}.
 
 In the approach of Refs.~\cite{Buchmueller:2014yoa,Malik:2014ggr} the messenger width was computed within the simplified
 model itself. But this assumes that the messenger can only decay into the DM particles as well as the $q\bar{q}$ pairs from which
 it was produced in the first place. 
  This is a strong assumption which we are not prepared to apply universally, as this would exclude
 the possibility of mediators decaying into anything except a single species of the cosmologically stable DM within the dark sector (and  would also limit the decay possibilities into SM particles). 
Instead, as already pointed out above, we will treat $\Gamma_{\rm MED}$ as a free parameter
 which we will vary and whose minimal value should not be less than the calculated width into DM and the appropriate SM channels\footnote{Recently while this paper was being finalised, Ref.~\cite{Buckley:2014fba} appeared, considering scalar and pseudo-scalar mediators in the gluon fusion channel and also pointing out the importance of keeping the mediator width a free parameter. Limits for invisible decays of the Higgs boson in mono-jet measurements, taking the full top-mass dependence into account, have been obtained in recent years \cite{Englert:2011us,Aad:2011xw}.}.
 We advocate the approach with four parameters: $m_{\rm MED}$, $\Gamma_{\rm MED}$, $m_{\rm DM}$ and the 
 product of the couplings of the mediator to the SM and to DM particles, $g^2_{\rm eff}$. In order to provide a fully flexible Monte Carlo tool for experimental studies we have implemented the models described above in the MCFM framework~\cite{Campbell:1999ah,Campbell:2011bn,MCFMweb}. The results of this paper extend the existing dark matter processes in MCFM~\cite{Fox:2012ru}, which focused primarily on NLO corrections in the effective field theory approach (matching to parton showers was achieved in ref.~\cite{Haisch:2013ata}.  The results we present here will be available in the next public release of MCFM. 
 
  \medskip
 
 This paper is organised as follows.
 We first briefly discuss the limitations of the EFT approach at collider searches
 and proceed to define and setup  the simplified models we study in Sec.~\ref{sec:model}.
 In Section~\ref{sec:DDandID} we assemble the necessary formulae for DM scattering cross sections for our models relevant to
 direct detection and indirect detection of DM experiments. 
 In Sec.~\ref{sec:searches} we discuss the event generation and reconstruction as well as existing measurements for mono-jet final states, and proceed
 to present limits and projections for our simplified models. 
 Following this, in Sec.~\ref{sec:nploop} we extend our simplified models to allow for the possibility of additional contributions of new very heavy particles to the mediator production mechanism from initial state gluons. In the Appendix for the convenience of the Reader
 we list the basic amplitudes for a scalar mediator plus jet production in the spinor helicity formalism.
 Our conclusions are presented in Sec.~\ref{sec:conclusion}.

\section{Simplified Models}
\label{sec:model}

In this article we focus primarily on mono-jet searches, induced by a mediator interaction between gluons and dark particles. We stress here that collider limits apply whether the dark particles are dark matter candidates or not, i.e. no assumption on their cosmic abundance or astrophysical production mechanism is required.

The promising kinematics of the mono-jet signature \cite{Aaltonen:2012jb, Chatrchyan:2012me, ATLASMONO, Khachatryan:2014rra,Diehl:2014dda} was appreciated for new physics searches several years ago \cite{Barger:1984uz, Glover:1984iu, Hall:1985wz, Beltran:2010ww}. Models studied include those with extra-dimensions \cite{Abazov:2003gp} to compressed SUSY spectra \cite{Hall:1985zn,Gunion:2001fu,Dreiner:2012gx}. More recently it was also argued that this configuration can constrain dark sectors in a fairly model-independent way \cite{Feng:2005gj,Cao:2009uw,Beltran:2010ww,Goodman:2010yf,Goodman:2010ku,Fox:2011pm,Haisch:2012kf}. To limit the number of free parameters and to facilitate the interpretation and cross correlation of measurements at colliders and direct detection experiments effective operators were proposed to parametrise the signal hypotheses, i.e. the contributions of new physics models.

\begin{center}
\begin{figure}
\includegraphics[width=0.6\textwidth]{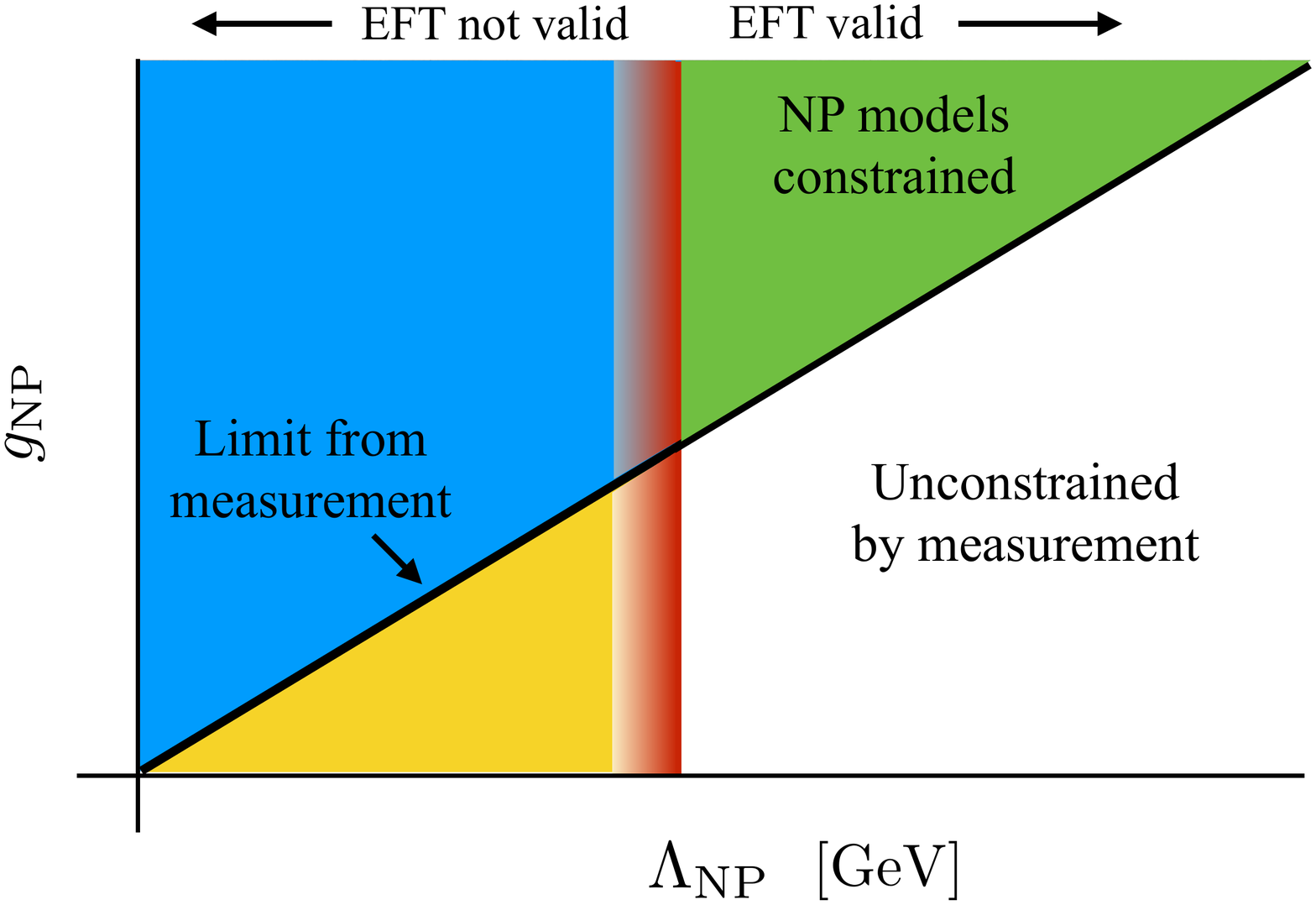}
\caption{Figure taken from \cite{Englert:2014cva}, schematically illustrating the valid interpretations of experimental results 
in terms of EFTs. Experimental analyses set a bound on the $g_{NP}/\Lambda$, corresponding the solid line, combinations of 
$g_{NP}$ and $\Lambda$ greater than this bound are excluded. However, if the experiment was able to probe the scale of the new 
physics then the EFT prescription was invalid, this corresponds to all values left of the vertical line.}
\label{fig:eft}
\end{figure}
\end{center}

A priori, following the discussion in \cite{Englert:2014cva}, the use of effective operators in constraining new physics scenarios in a fairly model independent way is a legitimate approach
in parts of the model parameter space.
Differential distributions can always be used to constrain the Wilson coefficients $C_i$ of specific effective operators $O_i$. 
It needs to be kept in mind, however, that these constraints are only meaningful when the scale at which the operators are probed is below the validity region of the effective theory, e.g. $\sqrt{\hat{s}} \ll \Lambda_{\mathrm{NP}}$. This constraint results in the red vertical line of Fig.~\ref{fig:eft}. Because $C_i \sim g_{\mathrm{NP}}/\Lambda_{\mathrm{NP}}$, a constraint from a measurement on the Wilson coefficient translates into a diagonal curve depicted in black in Fig.~\ref{fig:eft}, resulting in four regions of the parameter space of new physics models. While the sectors left of the vertical red line are outside the validity range of the effective theory, only the models that fall into the green region 
could be
constrained by the measurement. More specifically, when aiming for an interpretation of the constraint on the effective operator in terms of a new physics model, models that are constrained have to have a high new physics scale, i.e. $\sqrt{\hat{s}} < \Lambda_{\mathrm{NP}} \simeq m_{\mathrm{NP}}$, and a large coupling $g_{\mathrm{NP}}$. 
This can require the dark particles to be strongly coupled to the visible sector, which further complicates the interpretation.

Therefore,
a reliable interpretation of collider searches of dark matter particles should be based on basic QFT interactions 
where all intermediate propagating degrees of freedom in a given process are explicitly taken into account\cite{Busoni:2013lha,Papucci:2014iwa,Busoni:2014sya,Abdallah:2014hon,Haisch:2013fla} 
Unsurprisingly, contributions due to light degrees of freedom have been found to be significant for limit setting \cite{Fox:2011fx, Fox:2012ru, Buchmueller:2013dya,Alves:2013tqa,Arcadi:2013qia,Richard:2014vfa}. \\

In order to model mediator production, 
we will consider simplified models with the mediators to the dark sector associated with scalar $S$, pseudo-scalar $P$, vector $Z'$ and axial-vector
$Z''$ fields with interactions,
\begin{align}
\label{eq:LS} 
\mathcal{L}_{\mathrm{scalar}}&\supset\, -\,\frac{1}{2}m_{\rm MED}^2 S^2 - g_{\rm DM}  S \, \bar{\chi}\chi
 - g_{SM}^t S \, \bar{t} t  - g_{SM}^b S \, \bar{b} b\,,
 \\
 \label{eq:LP} 
\mathcal{L}_{\rm{pseudo-scalar}}&\supset\, -\,\frac{1}{2}m_{\rm MED}^2 P^2 - g_{\rm DM}  P \, \bar{\chi} \gamma^5\chi
 - g_{SM}^t  P \, \bar{t}  \gamma^5t  - g_{SM}^b  P \, \bar{b}  \gamma^5b\,,
 \\
 \label{eq:LV} 
\mathcal{L}_{\mathrm{vector}}&\supset \, \frac{1}{2}m_{\rm MED}^2 Z'_{\mu} Z'^{\mu} - g_{\rm DM}Z'_{\mu} \bar{\chi}\gamma^{\mu}\chi -\sum_q g_{SM}^q Z'_{\mu} \bar{q}\gamma^{\mu}q\,,
 \\
 \label{eq:LA} 
\mathcal{L}_{\rm{axial}}&\supset\,  \frac{1}{2}m_{\rm MED}^2 Z''_{\mu} Z''^{\mu} - g_{\rm DM} Z''_{\mu} \bar{\chi}\gamma^{\mu}\gamma^5\chi -\sum_q g_{SM}^q Z''_{\mu} \bar{q}\gamma^{\mu}\gamma^5q \,.
\end{align}
Two types of coupling constants appear in these equations: $g_{\rm SM}$ which collectively denote the couplings between messenger fields 
and Standard Model particles, and $g_{\rm DM}$ which are couplings of the messenger to the dark sector $\chi$ particles.
We have assumed that the scalar and pseudo-scalar messengers are coupled only to top and bottom quarks 
with the Yukawa-type coupling denoted $g_{SM}^{t,b}$ in Eqs.~\eqref{eq:LS}-\eqref{eq:LP} -- these are the dominant interactions
of (pseudo)-scalars with the SM fermions; in fact in most cases only the couplings to tops are important. 
Phenomenologically this resembles models with minimal flavour violation \cite{D'Ambrosio:2002ex} and a SM-Yukawa-like hierarchy for the mediator-fermion interactions.
The couplings of messengers to all six flavours of SM quarks are taken to be proportional to the corresponding Higgs Yukawa couplings, $y_q$,
and to make our definitions look symmetric we choose to parametrise the DM couplings in a similar fashion, so that,
\begin{equation}
{\rm for\, scalar \,\& \, pseudo-scalar\, messengers:} \quad g_{\rm SM}^q \equiv\,  g_q y_q\,, \quad  g_{\rm DM} \equiv\,  g_\chi y_\chi \,,
\quad{\rm where} \quad y_\chi \equiv\,\frac{m_\chi}{v}= \frac{m_{\rm DM}}{v}\,.
\label{eq:gdef}
\end{equation}
The product of the top and $\chi$ couplings to messengers is, 
\begin{equation}
 g_{\rm eff}^2 \,:=\, g_{\rm SM}^q \, g_{\rm DM}\,=\,  g_t g_\chi\,  y_t y_\chi \,=\, g_q g_\chi\,\frac{m_t m_{\rm DM}}{v^2}
\label{eq:geffSP}\,,
\end{equation}
and we keep the scaling $g_q$ flavour-universal for all quarks, so $g_t=g_q$.

All vectors and axial-vectors are assumed to be coupled to all quarks uniformly, hence the sums in 
Eqs.~\eqref{eq:LV}-\eqref{eq:LA} are over all quark flavours (with a universal gauge-type coupling denoted  $g_{\rm SM}$).
For the axial-vector and vector mediators we will use
\begin{equation}
g_{\rm eff}^2 \,:=\, g_{\rm SM} \, g_{\rm DM}\,.
\label{eq:geffVA}
\end{equation}

In our setup the Standard Model particles only interact via the mediator with the invisible sector, i.e. the particle $\chi$. Thus, all amplitudes contributing to the processes we will study in Secs.~\ref{sec:DDandID}-\ref{sec:nploop} are proportional to $g_{\rm eff}^2$ defined
in Eqs.~\eqref{eq:geffSP} and \eqref{eq:geffVA}.
 
 \medskip
 
It is important to stress that models derived from scalar and pseudo-scalar mediators provide some of the simplest realisations of a non-minimal Higgs sector in which the Standard Model Higgs 
interacts and can mix with the scalar mediators. 
Following the Higgs discovery there is a renewed interest in the literature in Higgs portal models 
where the scalar mediators are SM-singlets but the SM Higgs $h$ interacts with them via the interaction,
$\lambda_{\rm hp} |H|^2 |\Phi|^2$. 
The Higgs portal models with singlet scalar messengers will be treated in the same way as general scalar messengers.
These models provide a direct link with Higgs physics and also include 
theoretical scenarios which manifest a common origin of the
electroweak and the DM scales in Nature as was recently explored in 
Refs.~\cite{Englert:2013gz,Englert:2011yb,Khoze:2013uia,Hambye:2013dgv, Khoze:2014xha,Altmannshofer:2014vra,Heikinheimo:2013xua,Gabrielli:2013hma,Duerr:2013dza,Duerr:2013lka}.

More specifically, consider the case where the mediator is a complex scalar $\Phi$ which is a singlet
of the SM and interacts with it only via the portal interactions with the Higgs,
\begin{equation}
\label{eq:hp}
\mathcal{L}_{\rm portal}= \lambda_{\rm hp} |H|^2 |\Phi|^2 - g_{\rm DM}~\bar{\chi} \Phi \chi.
\end{equation}
Furthermore, we assume that $\Phi$ is charged under the gauge group of the dark sector and is coupled to other dark particles (which 
in \eqref{eq:hp} for simplicity are taken to be the dark fermions $\chi$ and $\bar\chi$, but this can be extended to include vector and scalar
dark particles). 
In models which contain no input mass scales in the microscopic Lagrangian, 
the vacuum expectation value for the field $\Phi$ can be generated quantum mechanically, 
e.g via the Coleman-Weinberg mechanism \cite{CW}
in the dark sector as explained in \cite{Englert:2013gz}. The VEV  $\left < \Phi \right >$ then induces the vacuum expectation value $v$ for the Higgs field via the portal interaction in Eq.~(\ref{eq:hp}) and triggers electroweak symmetry breaking. It also generates the 
mass scale $m_{DM} =  g_{\rm DM}~\left < \Phi \right >$ in the dark sector. Thus, in this class of models 
the electroweak scale and dark matter scale have a common origin.
To see that such Higgs portal models continue to be described effectively by the minimal simplified model in Eq.~(\ref{eq:LS})  
we re-write \eqref{eq:hp}
after electroweak symmetry breaking, in unitary gauge, as,
\begin{equation}
\mathcal{L}_{\rm portal} \supset 2 \lambda_{\rm hp} \left < \Phi \right > v~\phi h +  \lambda_{\rm hp} v^2~\phi^2
 +  \lambda_{\rm hp} \left < \Phi \right >^2 h^2 - g_{\rm DM}~\bar{\chi} (\left < \Phi \right >+\phi)  \chi.
 \end{equation}
Transforming into the mass eigenstate basis, we find two scalar resonances $h_1$ and $h_2$, both of which interact with the Standard Model and the dark sector. Either state can be identified with $S$ in Eq.~(\ref{eq:LS}).\footnote{For simplicity and concreteness, this paper 
concentrates on the simplified models with mediators coupled to Dirac fermions $\chi$ in the dark sector. These models can be extended
to incorporate scalar and vector dark matter particles as in \cite{Khoze:2013uia,Hambye:2013dgv, Khoze:2014xha} and chiral fermions.}

Following a similar line of reasoning for the case where the pseudo-scalar can develop a vev and CP is not a good quantum number we can map the Higgs portal interactions to the form of Eq.~(\ref{eq:LP}).

We note that the scalar and pseudo scalar Lagrangians defined above are compatible with the principle of minimal flavor violation. 
If on the other hand one were to relax this constraint one could define a scalar (and pseudo-scalar) Lagrangian in which the mediating 
particle couples directly to the light quark species, with no Yukawa suppression. In these instances the phenomenology of the signal 
changes substantially, since the production mechanism is now identical to vector and axial-vector mediators. As a result the phenomenology 
of these signatures, (LHC limits and cross sections) are similar in size to those obtained using the vector and axial mediators  (the major differences
arising from a scalar mediator is an isotropic final state with no spin correlations). We note that these processes are available in MCFM~\cite{Fox:2012ru}, and the analysis we present here could be applied easily to these models, however for brevity we do not consider them in this paper.

\subsection{The mediator width}
\label{sec:width} 

\begin{center}
\begin{figure}
\includegraphics[width=0.45\textwidth]{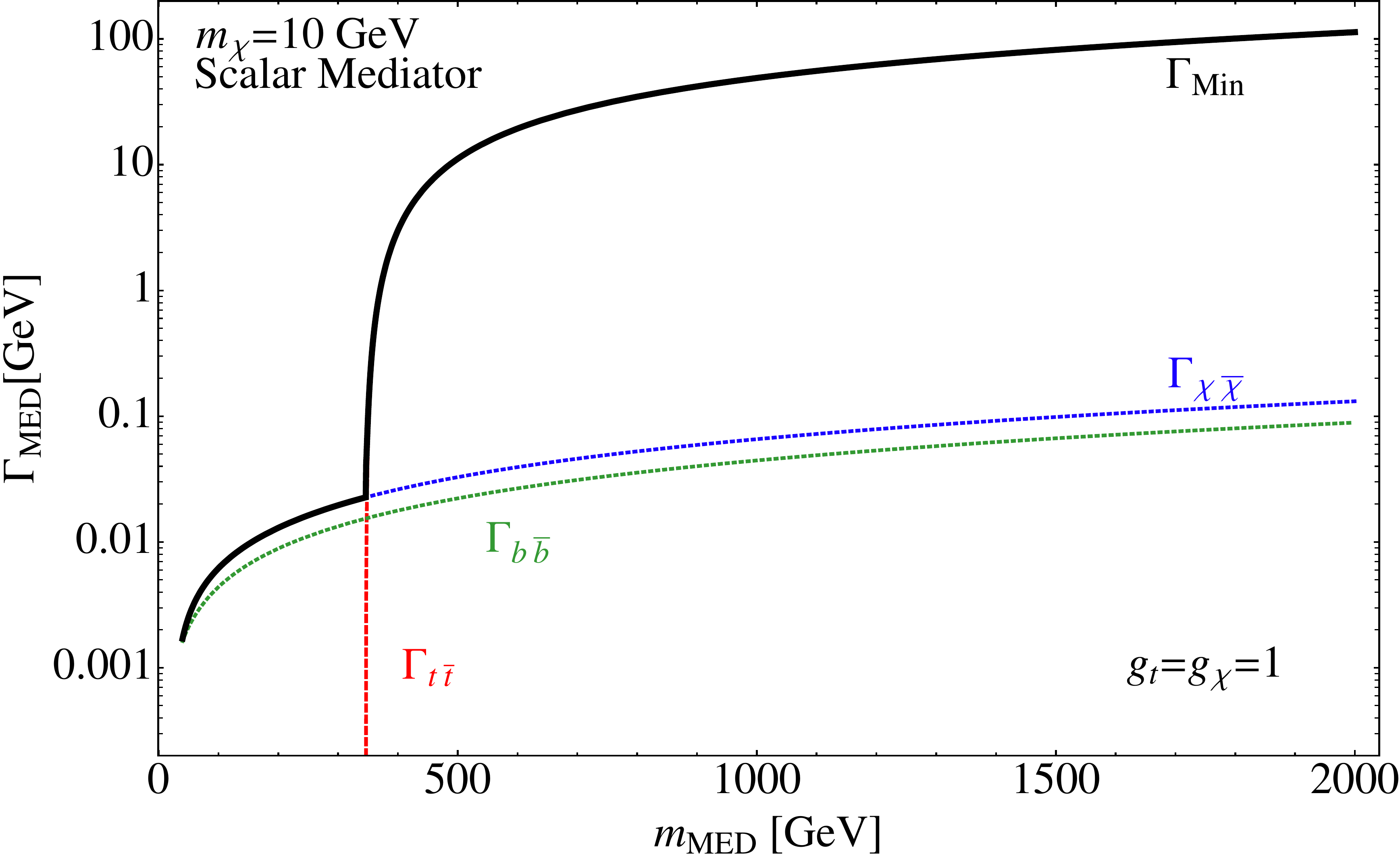}
\includegraphics[width=0.45\textwidth]{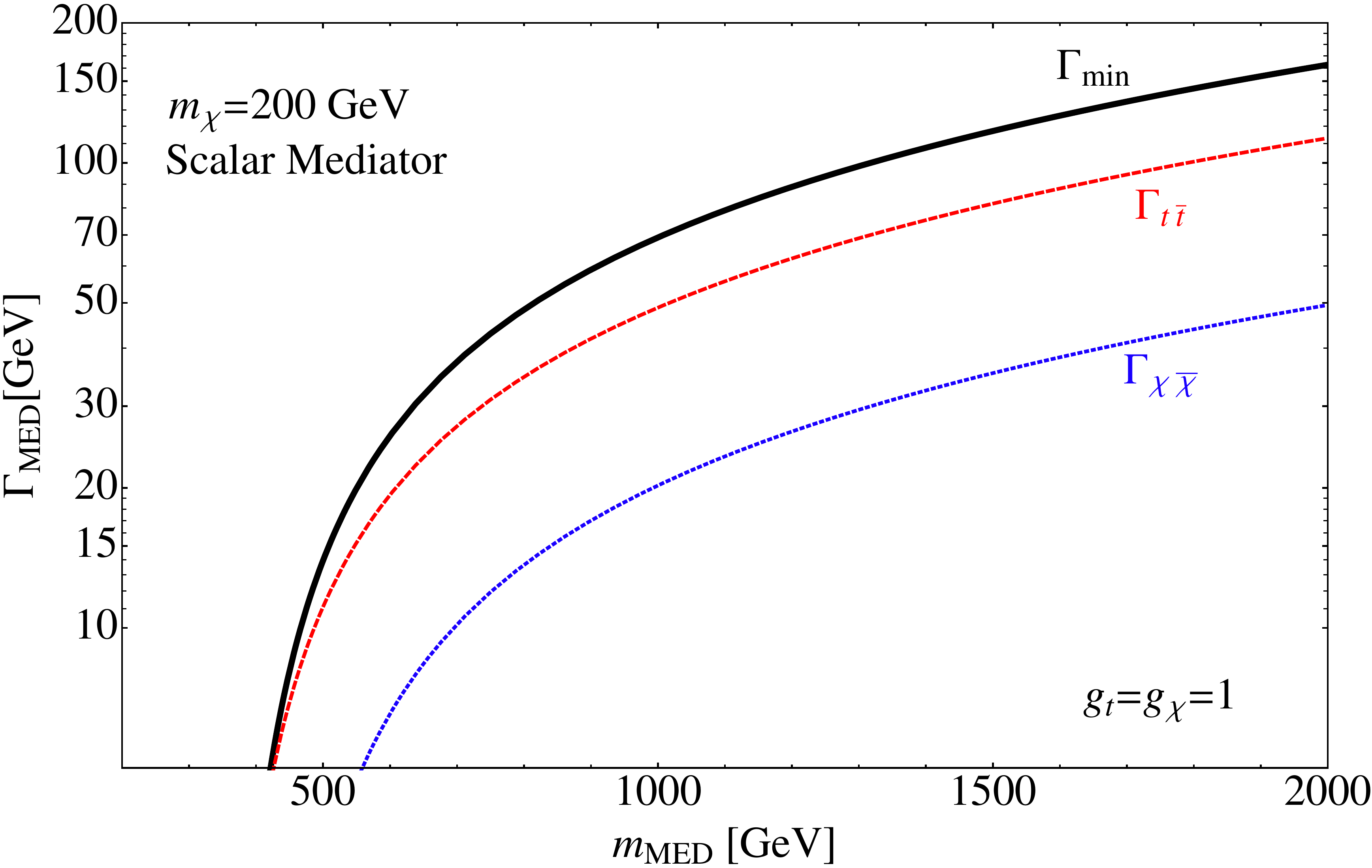}
\includegraphics[width=0.45\textwidth]{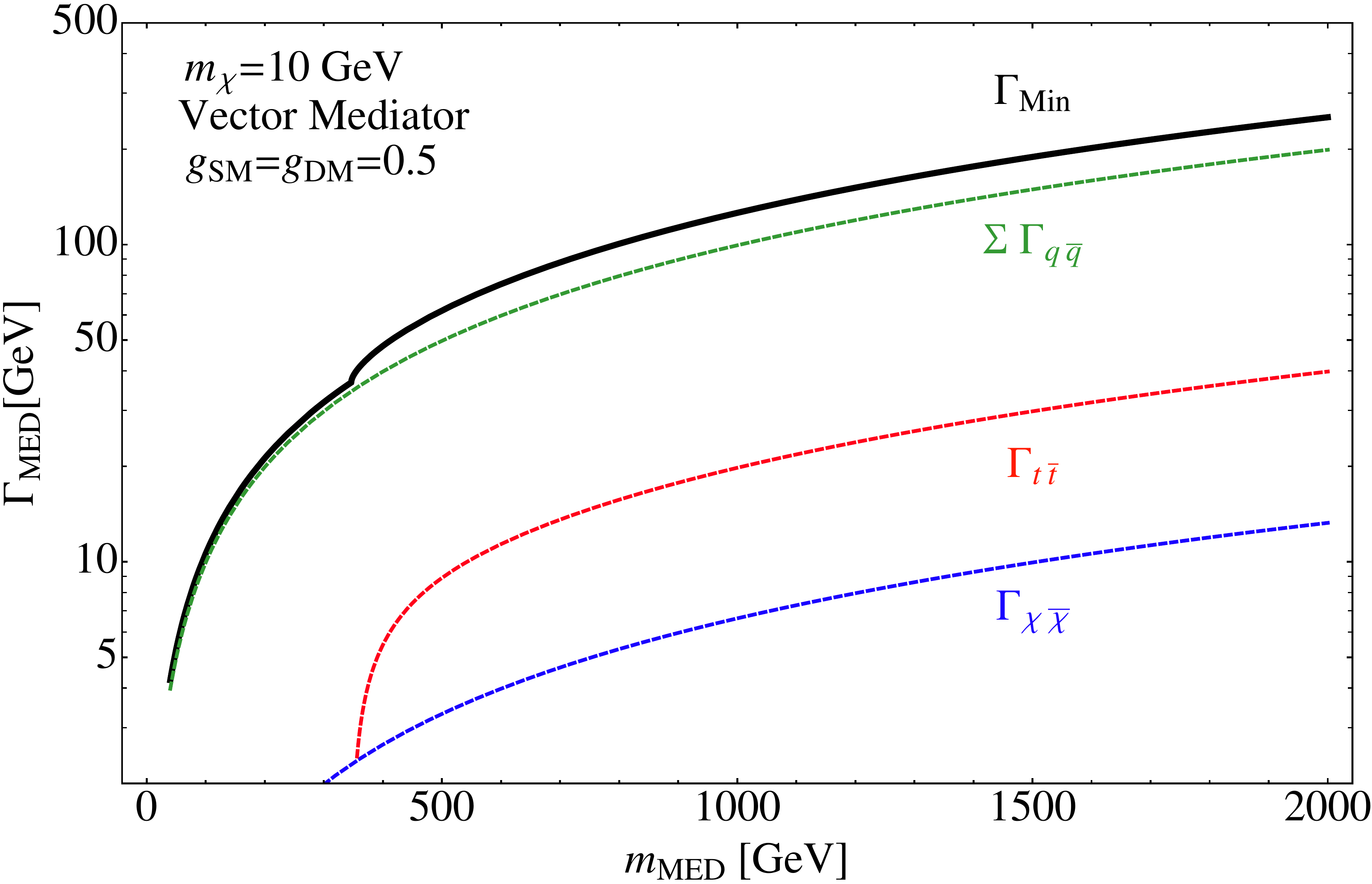}
\includegraphics[width=0.45\textwidth]{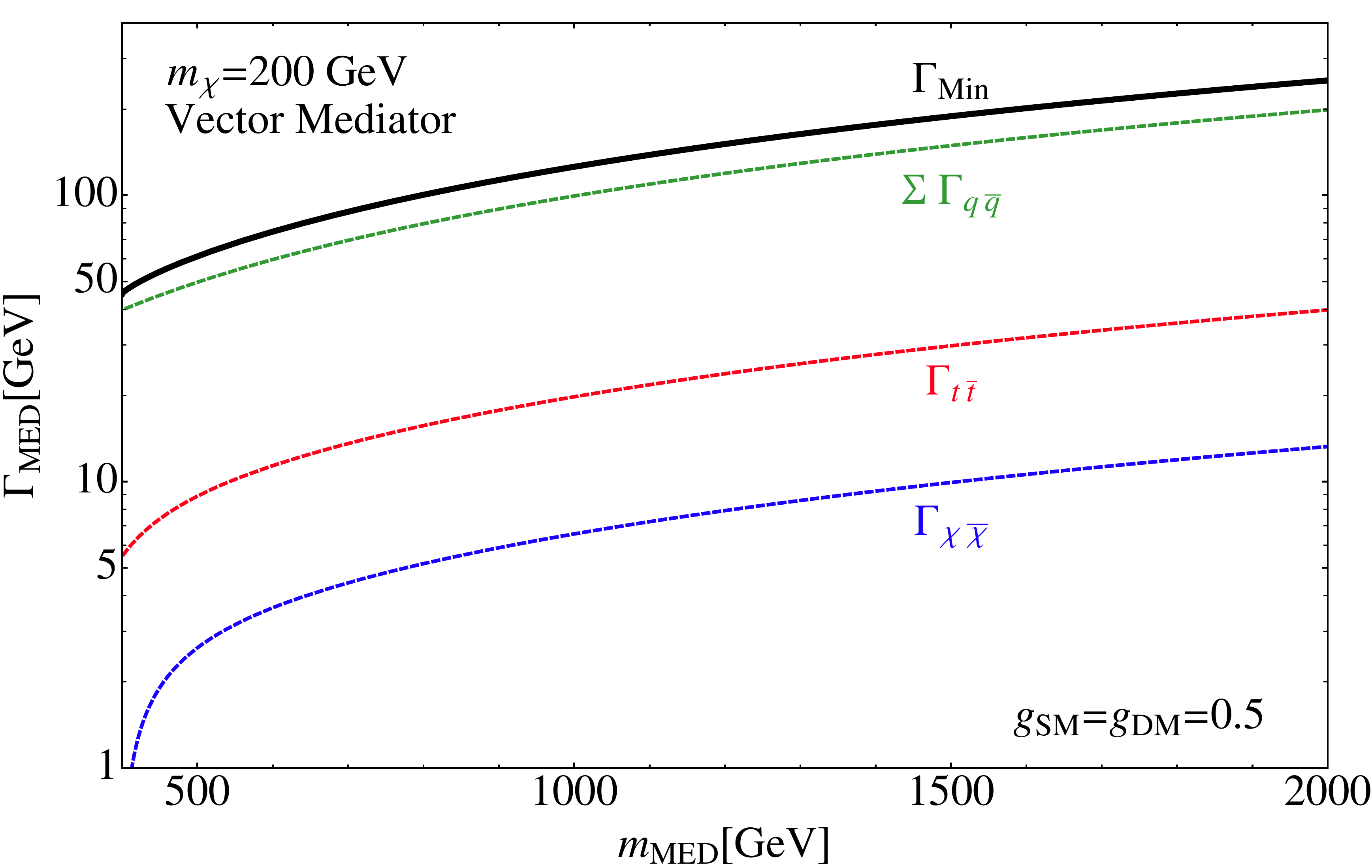}
\caption{Minimal width as a function of the mediator mass for scalar and vector operators, with two different DM mass choices. Individual partial width contributions are illustrated.}
\label{fig:min-width}
\end{figure}
\end{center}

We would now like to discuss the impact of the mediator width in our simplified models. Given the models specified in Eqs.~(\ref{eq:LS})-(\ref{eq:LA}) with democratic quark-(axial)vector and Yukawa-type quark-scalar interactions we obtain a lower limit for the width of the mediator. For scalar and pseudo-scalar mediators, depending on their mass, decays to heavy quarks may or may not be open (i.e. $m_{\rm MED}$ is required to be $> 2m_t$ for an open decay). In certain regions of parameter space, loop induced decays to vector bosons, or extended dark sector decays, and off-shell decays (e.g. to $t^*\overline{t}$), may significantly enhance the ``minimal widths" which we define as, 
\begin{eqnarray}
\label{eq:GVA}
\Gamma^{V,A}_{\rm MED, min}&=&\Gamma^{V,A}_{\chi\overline{\chi}} + \sum_{i=1}^{N_f} N_c \Gamma^{V,A}_{q_i\overline{q}_i} + N_c \Gamma^{V,A}_{t\overline{t}}  \\
\label{eq:GSP}
\Gamma^{S,P}_{\rm MED, min}&=&\Gamma^{S,P}_{\chi\overline{\chi}} + N_c \Gamma^{S,P}_{t\overline{t}} 
\end{eqnarray}
where $\Gamma_{\chi\overline{\chi}}$ is the mediator decay rate into two DM particles (which here
we assume are fermions $\chi\overline{\chi}$, modifications to scalar dark matter are trivial to incorporate). The sum on the right hand side of the first equation is over the massless SM quark flavours interacting with the vector and axial-vector mediators.
These widths are lower bounds on the total and as such we treat the width as a free parameter and investigate 
the LHC phenomenology as a function of the rescaled width. For decays into fermions the partial widths are defined as follows, 
\begin{eqnarray}
\label{eq:GV}
\Gamma^{V}_{f\overline{f}} &=& \frac{g^2_{f}(m_{\rm MED}^2+2m_f^2)}{12 \pi m_{\rm MED}}\sqrt{1-\frac{4m_f^2}{m_{\rm MED}^2}} \\
\label{eq:GA}
\Gamma^{A}_{f\overline{f}} &=& \frac{g^2_{f}(m_{\rm MED}^2-4m_f^2)}{12 \pi m_{\rm MED}}\sqrt{1-\frac{4m_f^2}{m_{\rm MED}^2}} \\
\label{eq:GS}
\Gamma^{S}_{f\overline{f}} &=& \frac{g^2_{f} m_f^2m_{\rm MED}}{8\pi v^2}\left(1-\frac{4m_f^2}{m_{\rm MED}^2}\right)^\frac{3}{2}\\
\label{eq:GP}
\Gamma^{P}_{f\overline{f}} &=& \frac{g^2_{f} m_f^2 m_{\rm MED}}{8\pi v^2}\left(1-\frac{4m_f^2}{m_{\rm MED}^2}\right)^\frac{1}{2}
\end{eqnarray}
where $m_f$ denotes masses of either SM quarks $q$ or DM fermions $\chi$ and the coupling constant $g_{f}$ denotes
either $g_q$ or $g_\chi$ as defined on the right hand side of Eq.~\eqref{eq:gdef}.
In Fig.~\ref{fig:min-width} we plot the minimal widths computed using Eqs.~\eqref{eq:GVA}-\eqref{eq:GP} for scalar and vector types of the
mediators as functions of the mediator mass, for two representative choices of DM masses.
 As expected, the (pseudo)-scalar models parameterised 
in terms of Yukawa couplings, are much more sensitive to the choice of DM mass. The hadronic branching ratio for the vector mediator dominates 
the decays (due to the combination of light flavours and color factors $N_fN_c$), extended darks sectors could result in larger branching ratios to the dark sector 
and thus increase the width. For the scalar there are no light decays (apart from $b\overline{b}$ which can become important for light mediators), and the relative enhancement/suppression of $t\overline{t}$ decays scales like $ N_c(m_t/m_{DM})^2$. 

\medskip

\noindent{\bf In summary:} The set of simplified models for dark particle searches we study is defined by Eqs.~\eqref{eq:LS}-\eqref{eq:LA}.
It automatically accounts for Higgs portal models with scalar and pseudo-scalar messengers.
In Sec.~\ref{sec:nploop}
we will further extend the model in Eq.~\eqref{eq:LS} by adding a new BSM interaction \eqref{eq:eftLS}.

Our simplified models are characterised 
by the type of the mediator field, which can be a scalar, pseudo-scalar, vector or axial-vector particle.
There are four (five) types of input parameters involved in this description: the mediator mass $m_{\rm DM}$, the mediator width 
$\Gamma_{\rm DM}$, the dark particle mass $m_{\rm DM}$ and an appropriately defined coupling constant (or constants) to
characterise the combined effect of the SM-mediator and the mediator-Dark sector interaction.

We use $g_{\rm SM} \, g_{\rm DM}$ as the input effective coupling parameter for the vector
and axial-vector cases \eqref{eq:LV}-\eqref{eq:LA}. In the cases of scalar and pseudo-scalar mediators \eqref{eq:LS}-\eqref{eq:LP}
we choose to scale the couplings with the SM Yukawa's and use the product of scaling factors $g_q \, g_\chi$ defined in \eqref{eq:gdef}
as the input effective coupling parameter.\footnote{The choice of what is treated as the input coupling parameter for
(pseudo)-scalar mediators, namely the combination 
 $(g_{\rm SM}/y_q)(g_{\rm DM}/y_\chi)$, or  $(g_{\rm SM}/y_q) g_{\rm DM}$, or the original couplings $g_{\rm SM} \, g_{\rm DM}$
 is of course only a simple re-parametrisation which only affects which dimensionless parameters are held fixed when one varies 
the mass parameters for the mediators/dark particles in the plots. We have chosen the first combination, the authors of 
\cite{Buckley:2014fba} used the second, while the vector cases of course have no Yukawa's to scale. In any case this is a simple
re-parametrisation.} (The extended model studied in Sec.~\ref{sec:nploop} will contain an additional coupling $g_g$.)

\section{Direct and Indirect Detection limits}
\label{sec:DDandID}

If we make the assumption that the particle $\chi$ of Eqs.~(\ref{eq:LS})-(\ref{eq:LA}) is a dark matter candidate, accounting for the observed dark matter abundance in the universe, we can derive limits on our simplified models from low-energy interactions, i.e. direct and indirect detection experiments. 
Direct detection experiments measure the recoil of the nucleus of which the dark matter particle scatters off. Our limits are based on measurements by LUX \cite{Akerib:2012ys,Akerib:2013tjd,Szydagis:2014xog} which currently provides the strongest bounds for $m_{\rm DM} \gtrsim 6$ GeV.
In these settings dark matter particles are assumed to be non-relativistic, the momentum transfer (depicted in the right diagram of Fig.~\ref{fig:feyn}) is small and describing the interaction in terms of effective operators is justified as long as $\mathcal{O}(m_{\rm MED}) \gtrsim 1$ GeV.

For the calculation of the scattering cross section of a dark matter particle scattering spin-independently via a vector mediator off a proton we find
\begin{equation}
\sigma_{\mathrm{\chi p}}^V = \frac{9}{\pi} \frac{g_{\rm DM}^2 g_{\rm SM}^2 \rho^2}{m^4_{\mathrm{MED}}}
\label{eq:16}
\end{equation}
and for the scalar, interacting with the nuclei only via the gluons, we use \cite{Kurylov:2003ra,Hisano:2010ct, Cheung:2013pfa}
\begin{equation}
\sigma_{\mathrm{\chi p}}^S =  \frac{\rho^2}{\pi}  \left | \frac{m_p }{m_t} \frac{g_{t}y_t~g_{\chi}y_\chi}{m^2_{\mathrm{MED}}} \frac{2}{27} f_{\mathrm{TG}} \right |^2,
\label{eq:17}
\end{equation}
where $\rho =m_{{\rm DM}} m_{p}/(m_{\rm DM} + m_{p})$ is the reduced mass and $f_\mathrm{TG}\simeq 0.9$ \cite{Cheng:2012qr}.

Axial-vector mediators result in spin-dependent dark matter-proton scatterings, with the cross section described by  \cite{Dienes:2013xya}
\begin{equation}
\sigma_{\mathrm{\chi p}}^A = \frac{3}{\pi}\frac{g_{\rm DM}^2 g_{\rm SM}^2 a^2 \rho^2}{m^4_{\mathrm{MED}}},
\label{eq:18}
\end{equation}
with $a= \Delta u + \Delta d + \Delta s \simeq 0.43$ \cite{Cheng:2012qr, Buchmueller:2013dya}, assuming democratic couplings to all quark flavors.
We compare the predicted cross sections with the combined bounds of Wimp-proton-scattering limits of PICASSO \cite{Archambault:2012pm}, COUPP \cite{Behnke:2012ys} and SIMPLE \cite{Felizardo:2011uw}.

While direct detection experiments can give strong constraints for the vectors and the scalar mediator, the scattering of a pseudo-scalar off a nucleus is strongly velocity dependent and vanishes in the non-relativistic limit. Therefore, for pseudo-scalars, taking existing limits into account \cite{Ackermann:2011wa,Abdo:2010ex}, indirect detection experiments can result in stronger bounds than direct detection experiments \cite{Zheng:2010js, Boehm:2014hva}. For the simplified model of Eq.~(\ref{eq:LP}), using the s-wave velocity-averaged DM annihilation cross section into $\bar{b}b$,
\begin{equation}
\left < \sigma v \right >_{\bar{b}b}^P = \frac{N_C}{2 \pi} \frac{(y_b g_{b})^2 (y_\chi g_{\chi})^2  \, m_{\rm DM}^2 } {(m_{\rm MED}^2 - 4 m_{\rm DM}^2)^2 + m_{\rm MED}^2 \Gamma_{\rm MED}^2} \sqrt{1 - \frac{m_b^2}{m^2_{DM}}},
\label{eq:19}
\end{equation}
we can derive a bound on the parameters in the $\bar{b}b$ channel \cite{Ackermann:2011wa}. As always, the scaling factors $g_q$ 
are kept flavour-universal, so $g_{b}=g_t= g_q$ in these equations.

In all four expressions for the cross sections \eqref{eq:16}-\eqref{eq:19} for scalar and pseudo-scalar mediators we will keep $g_q$ and $g_\chi$ fixed, while for vector and axial-vector mediators we 
fix $g_{\rm SM}$ and $g_{\rm DM}$.

\section{Searches at the LHC}
\label{sec:searches}

 \subsection{Event generation and final state reconstruction}
 \label{sec:event}

\begin{center}
\begin{figure}[t]
\includegraphics[width=0.45\textwidth]{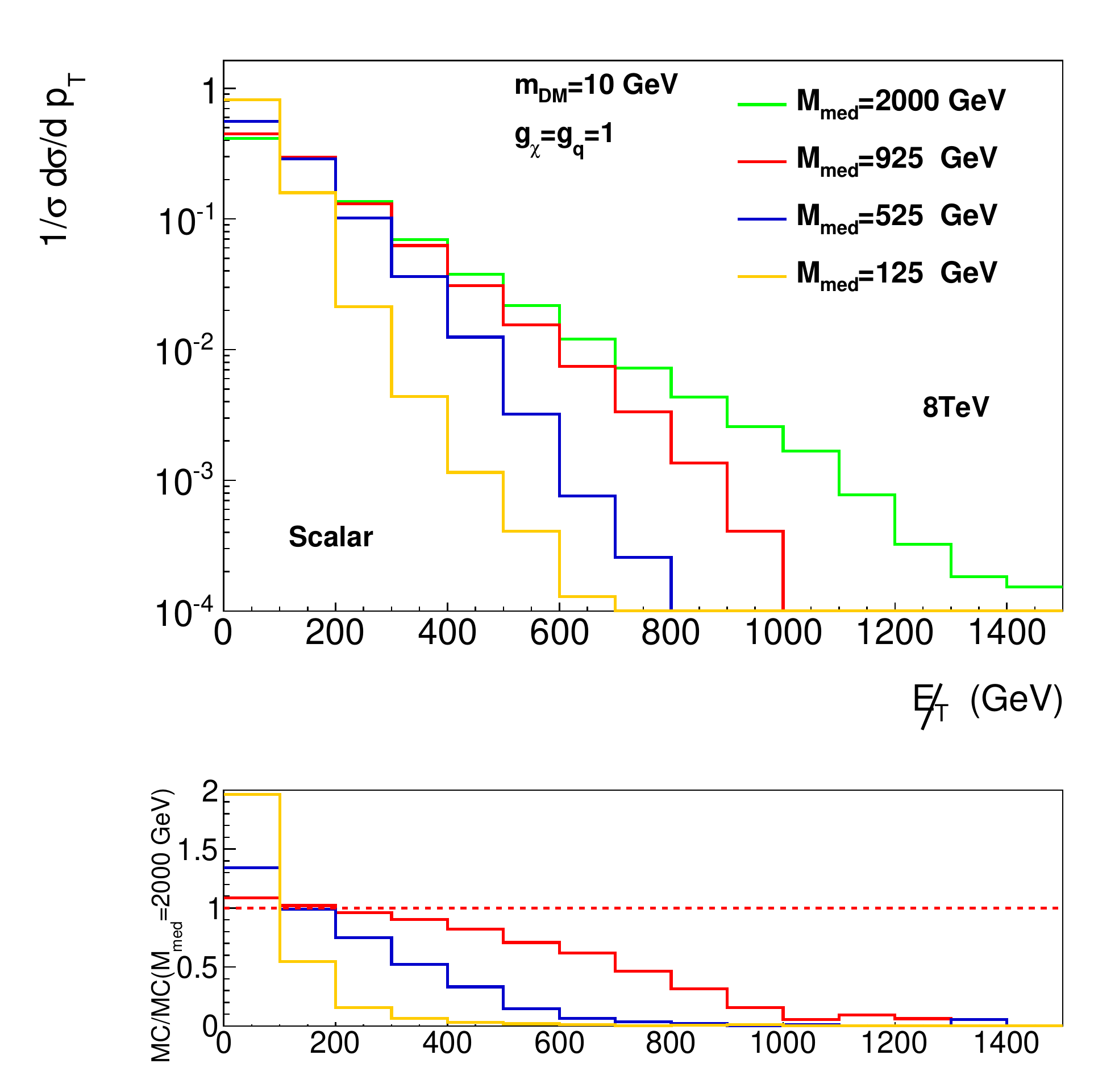}
\includegraphics[width=0.45\textwidth]{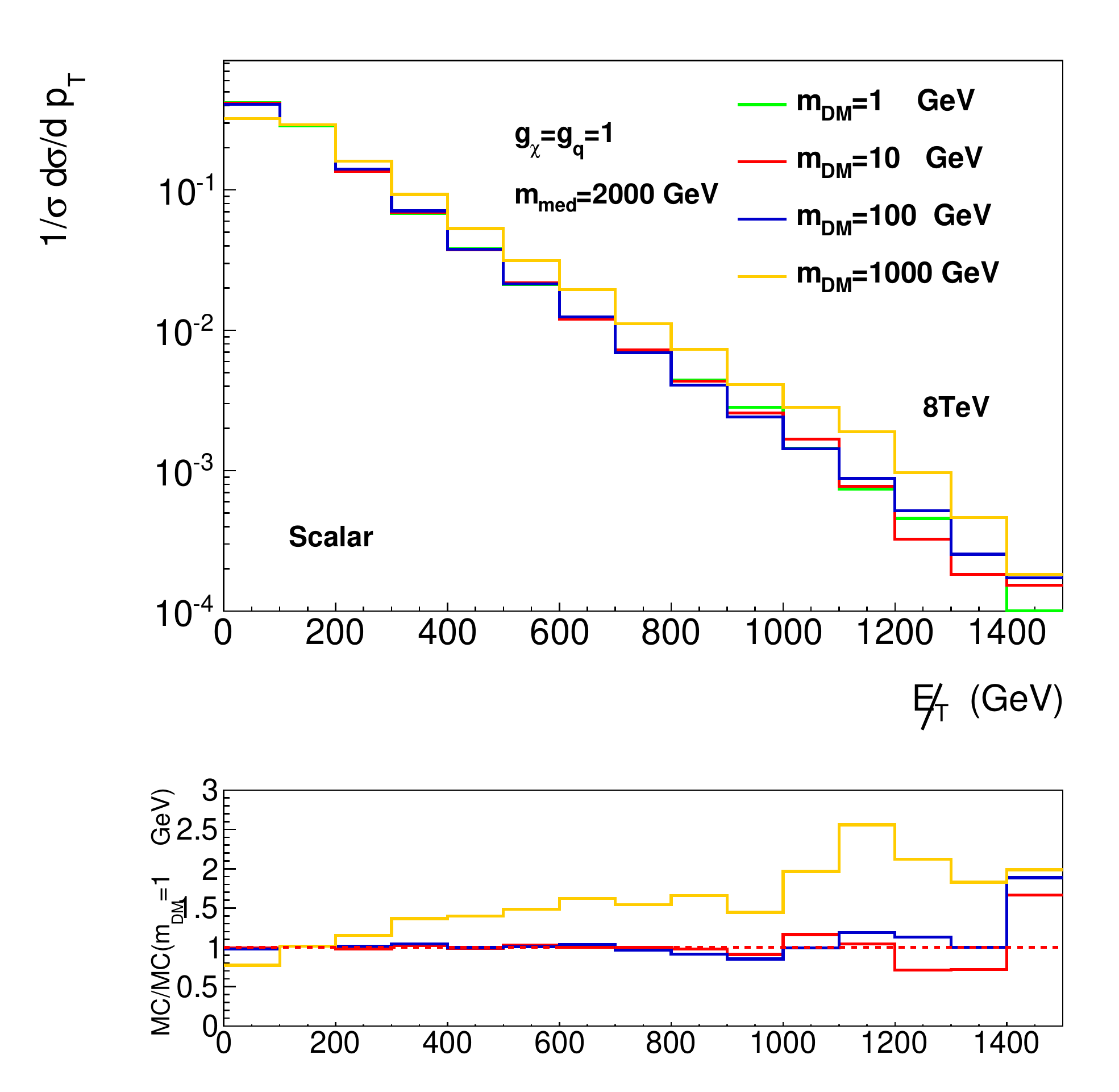} \\ \vspace{1.0cm}
\includegraphics[width=0.45\textwidth]{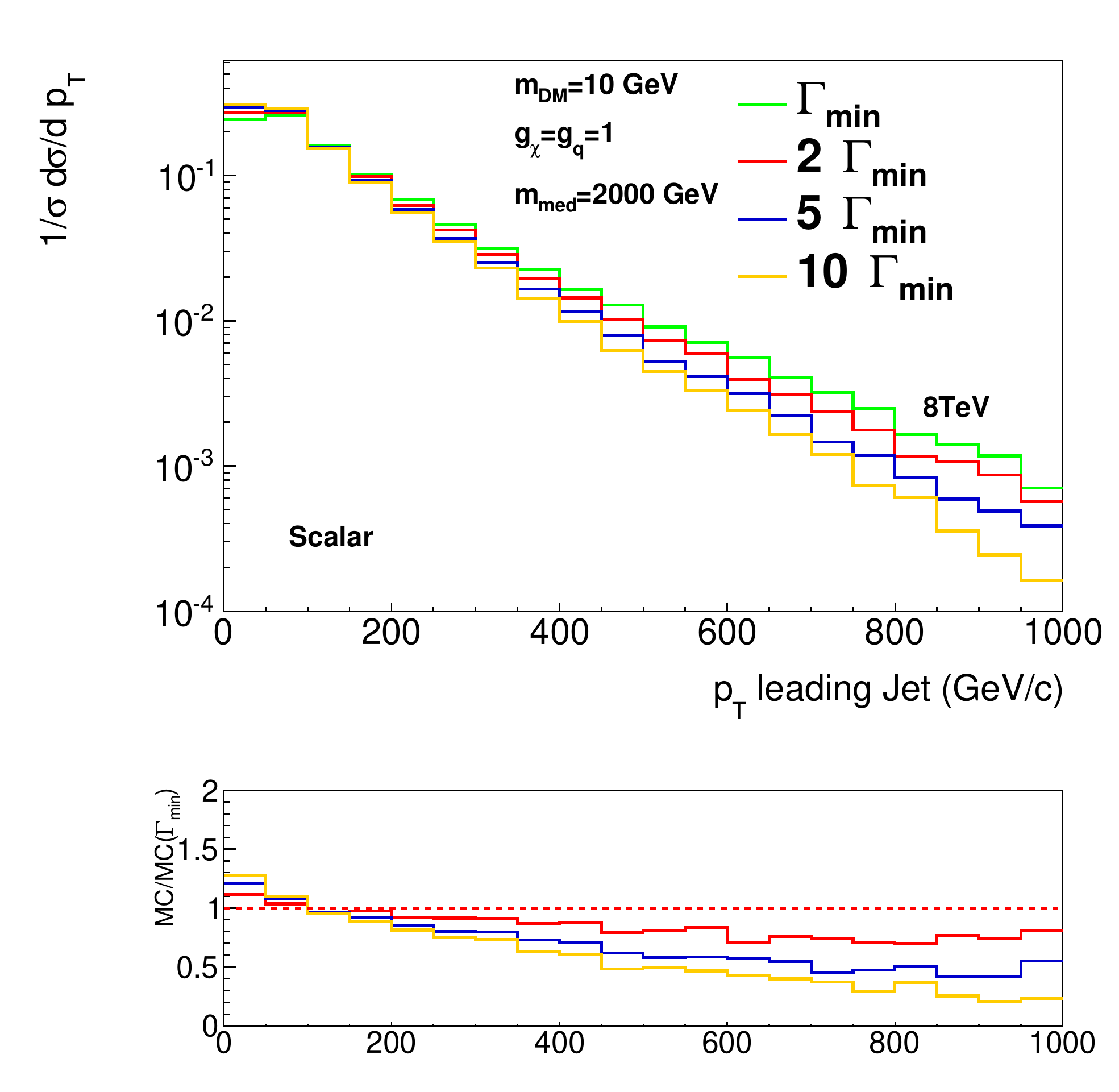}
\includegraphics[width=0.45\textwidth]{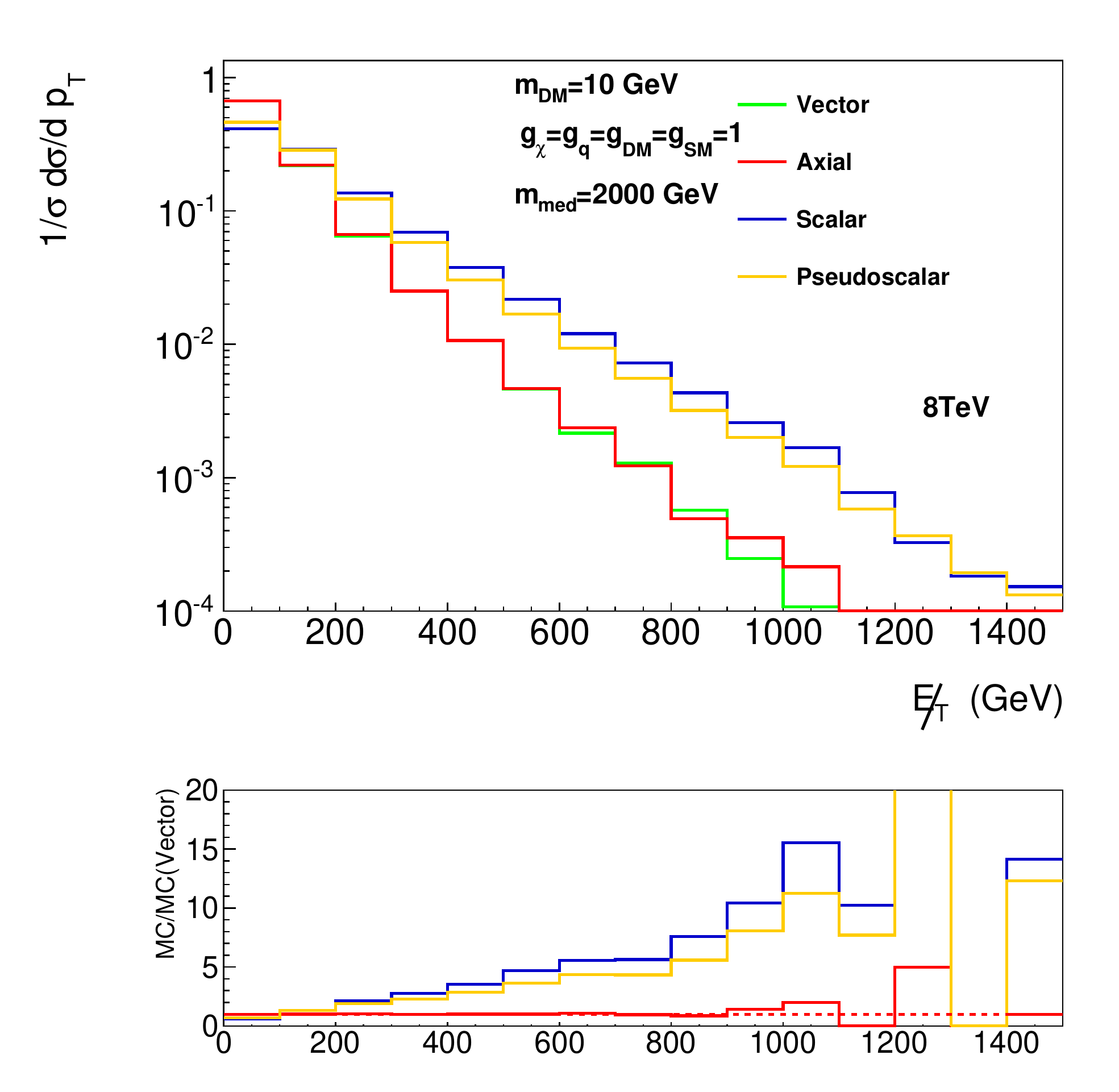}
\caption{ Kinematics and reconstruction of selected distributions. Events are generated
  using MCFM and PYTHIA, detector level effects are also simulated.
   Plots are normalised to have the same yield so that
  kinematic differences can be observed. 
We set the coupling parameters defined in Eq.~\eqref{eq:gdef} to $g_\chi=1=g_q$ for (pseudo)-scalars and
$g_{\rm DM}=1=g_{\rm SM}$ for (axial)-vector mediators.}
\label{fig:kinematics}
\end{figure}
\end{center}

We generate the processes $pp \to (X \to \bar \chi \chi) + {\rm jet}$
where $X$ can be a vector, axial-vector, scalar or pseudo-scalar using
MCFM. The existing implementation of DM process in MCFM (described in ref.~\cite{Fox:2012ru})
consisted of NLO predictions for vector, axial-vector, scalar and pseudo-scalar mediators in the effective 
field theory prescription, in addition LO processes involving a scalar mediator and top quark loop were also included. 
In the previous version propagating resonances were included for the above process in an {\it{ad hoc}} manner 
in which the user specified the value of the width, mediator mass and couplings, in addition to the mass of the dark matter particle. 
We have extended the code to include the LO process in which a pseudo scalar mediator couples to the top quark loop, and 
parameterized the code in terms of the simplified models described in the previous section. In addition we have included 
heavy new physics in the loop for scalar and pseudo-scalar mediator, but we postpone discussion of these effects to the next section. 
These extensions to the code will be released publicly in the next version of MCFM.  

 The generated signal samples are showered using Pythia 8
\cite{Sjostrand:2014zea} with tune 4C.  The background yield for
our 8 TeV limits are entirely inferred from the CMS mono-jet searches
\cite{Khachatryan:2014rra}.  To emulate the detector performance the
showered samples are clustered using the anti-$k_T$~\cite{Cacciari:2011ma}
 algorithm with a cone size of 0.4. The resulting three leading jets
with a $p_T$ above 15 GeV are smeared following the resolution
functions quoted by CMS~\cite{JET-CMS}. The remaining hadronic recoil is
smeared following the MET resolution quoted in \cite{MET-CMS}. 

For the 14 TeV analysis, the two leading backgrounds ($Z\to\nu\nu$ and
$W\to\ell\nu$ ) are generated using MadGraph \cite{Alwall:2011uj} and are
smeared according to the same scheme discussed in the previous
paragraph. From these samples a kinematic scale factor is obtained. 
The cross-sections for all other process, excluding the $t\overline{b}$ and
single top, are obtained by scaling the 8~TeV predictions in the CMS analysis by the
 NLO scale factor from 8~TeV to 14~TeV obtained in MCFM \cite{Campbell:2011bn}. For the top and
$t\overline{t}$ backgrounds the partial NNLO cross sections are used \cite{Kidonakis:2010dk}.

To be able to re-interpret the CMS mono-jet search in terms of the
simplified models of Eqs.~(\ref{eq:LS})-(\ref{eq:LA}) we follow the
event selection of \cite{Khachatryan:2014rra} closely. The signal
extraction region requires a $E_{\rm T}^{\rm miss} > 200$~\GeV, which
is beyond the plateau efficiency and thus, very  close to 100\%. The
hardest jet in the event must fulfill $p_{T,j_1} \geq 110$ GeV and
$|\eta| < 2.4$. We accept events with a second jet if $p_{T,j_2} \geq
30$ GeV, $|\eta| < 5.0$ and provided the azimuthal angle between the
two jets is less than 1.8 radiants, i.e. $\Delta \phi_{j_1,j_2} <
1.8$. We veto events with more than two jets. To optimise the
sensitivity of the analysis with respect to varying dark matter and
mediator masses, the exclusion limits are based on 7 phase space
regions distinguished by $E_{\rm T}^{\rm miss} >
(250,~300,~350,~400,~450,~500,~550)$ GeV.  

The limits shown in this section are based on the most
sensitive of the 7 regions respectively. For each signal model, the 7
regions are scanned taking the most sensitive result obtained by
scanning the individual regions. The limit computation is preformed in
the same manner as that in the CMS analysis, profiling the likelihood and obtaining
the 90\% confidence level using the CL$_{S}$ procedure. Both the
systematics and statistical uncertainties as described in the CMS
mono-jet analysis are taken into account. As a cross check, the cross
section limit results were reproduced to within 3\%.
For $14$ TeV we use the same reconstruction approach, keeping the same
systematics, but scaling the yields by the predicted scale factors.

In Fig.~\ref{fig:kinematics} we present a series of differential distributions obtained using the prescription defined above, focusing 
primarily on the case of a  scalar mediator. Most of the kinematic properties can be inferred from the on-shell condition for the mediator, namely 
\begin{eqnarray}
(\slashed{E}^{min}_T)^{2} + 4 m_{DM}^2 < s_{\chi\overline{\chi}}  \sim M_X^2  
\end{eqnarray}
As the mediator particle becomes heavier the missing transverse energy spectrum becomes harder, as illustrated by the top left plot 
in Fig.~\ref{fig:kinematics}. The lower left plot in Fig.~\ref{fig:kinematics} illustrates the jet $p_T$ dependence on the width. Broader widths result 
in a spreading of $s_{\chi\overline{\chi}}$, however events which become more energetic are additionally damped by PDF suppression. Therefore 
as particles acquire broader widths distributions are naturally softened relative to the narrow width case. Finally the lower right plot in Fig.~\ref{fig:kinematics}
illustrates the differences in the MET spectrum associated with the
production mechanism, the gluon induced scalar and pseudo scalar
processes result in a harder spectrum. 

Small kinematic differences are observed between each of the gluon
induced processes or vector induced processes. The largest modifications to the kinematic shape come from variations in the
mediator mass, and variation in the width. When scanning the dark matter
mass, visible modifications of the kinematics shapes are only present
for off-shell masses.

\subsection{Experimental Searches}
\label{sec:exp}

\begin{center}
\begin{figure}[h]
\includegraphics[width=0.47\textwidth]{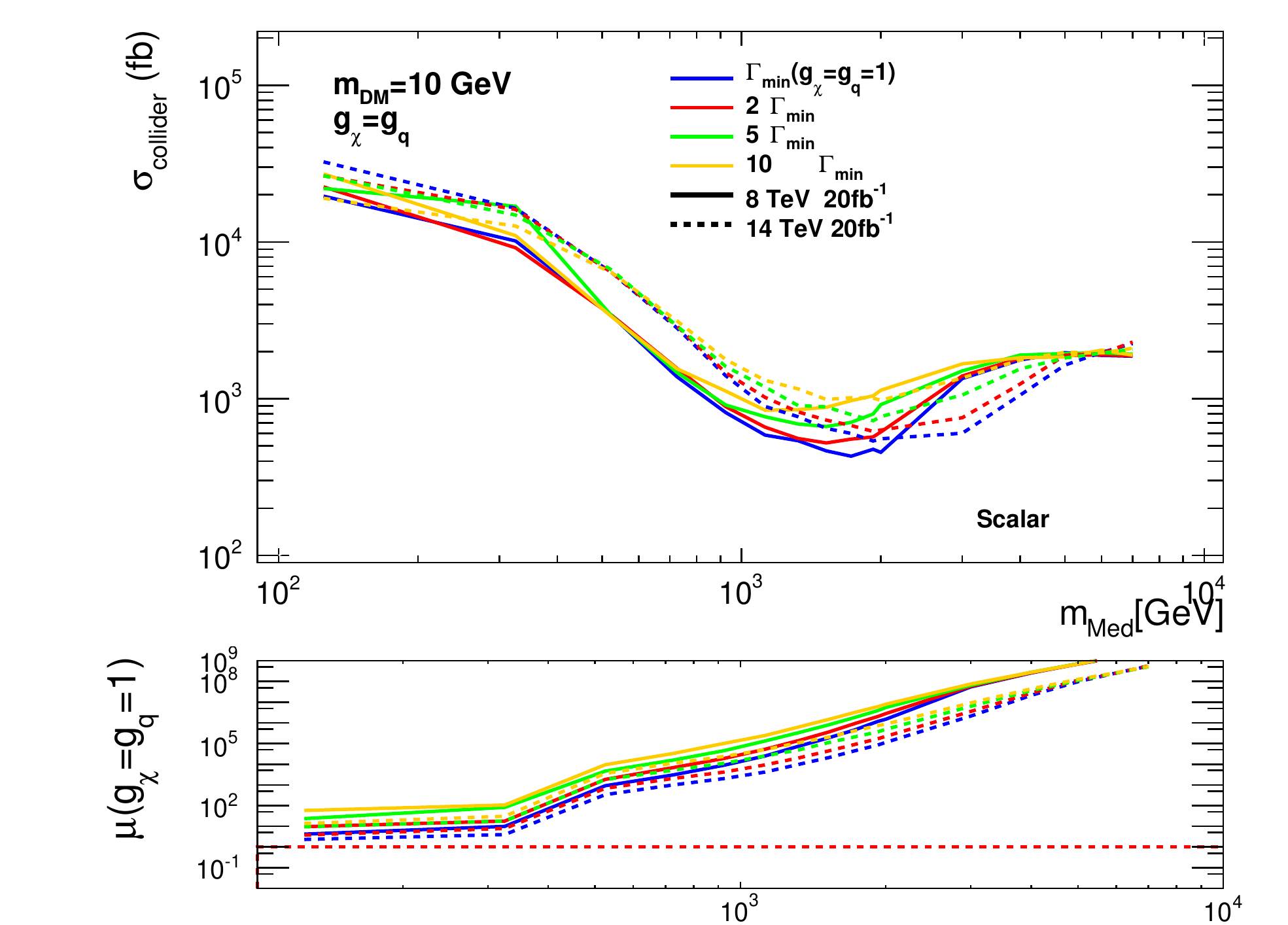}
\includegraphics[width=0.47\textwidth]{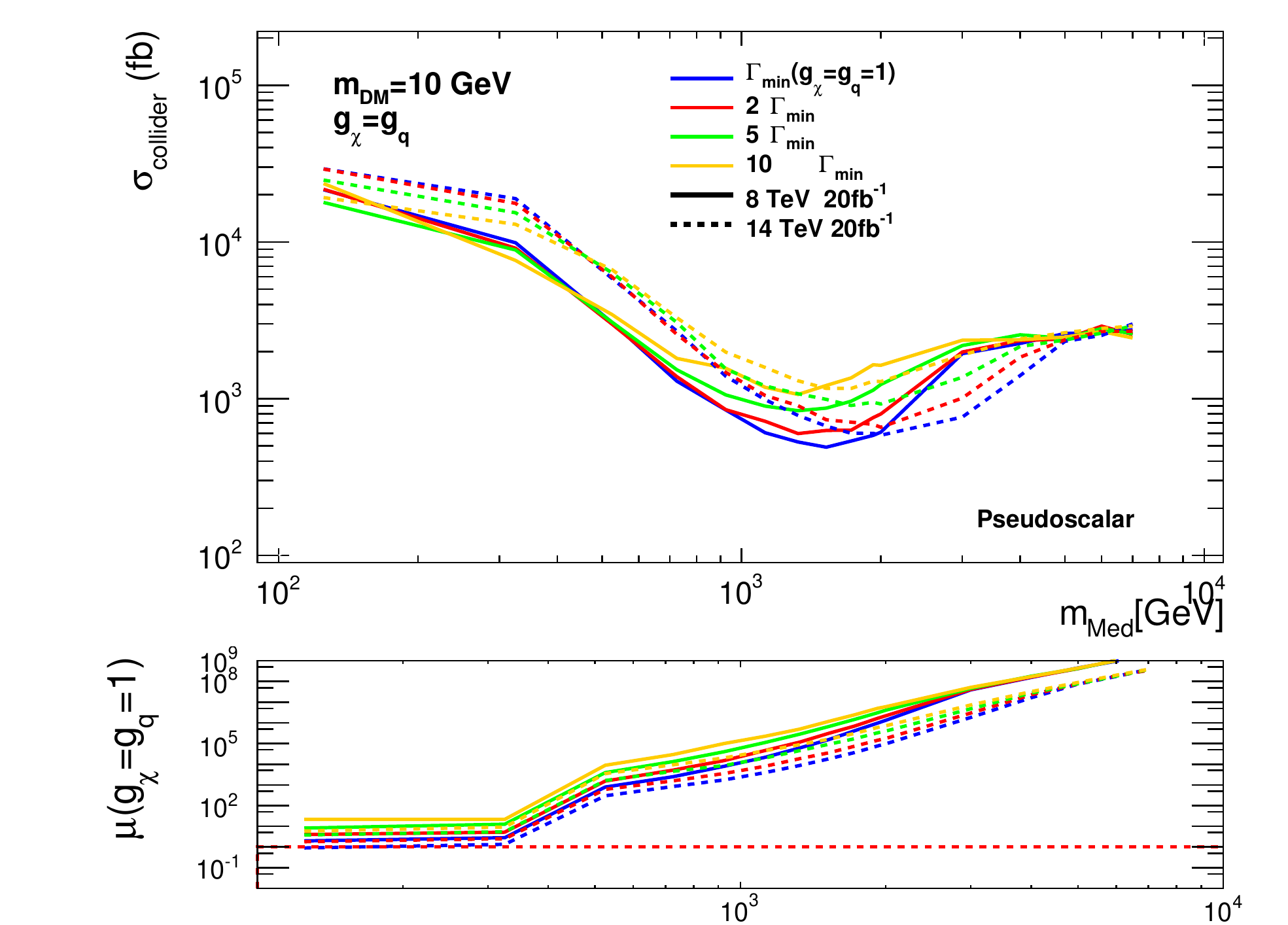}
\includegraphics[width=0.47\textwidth]{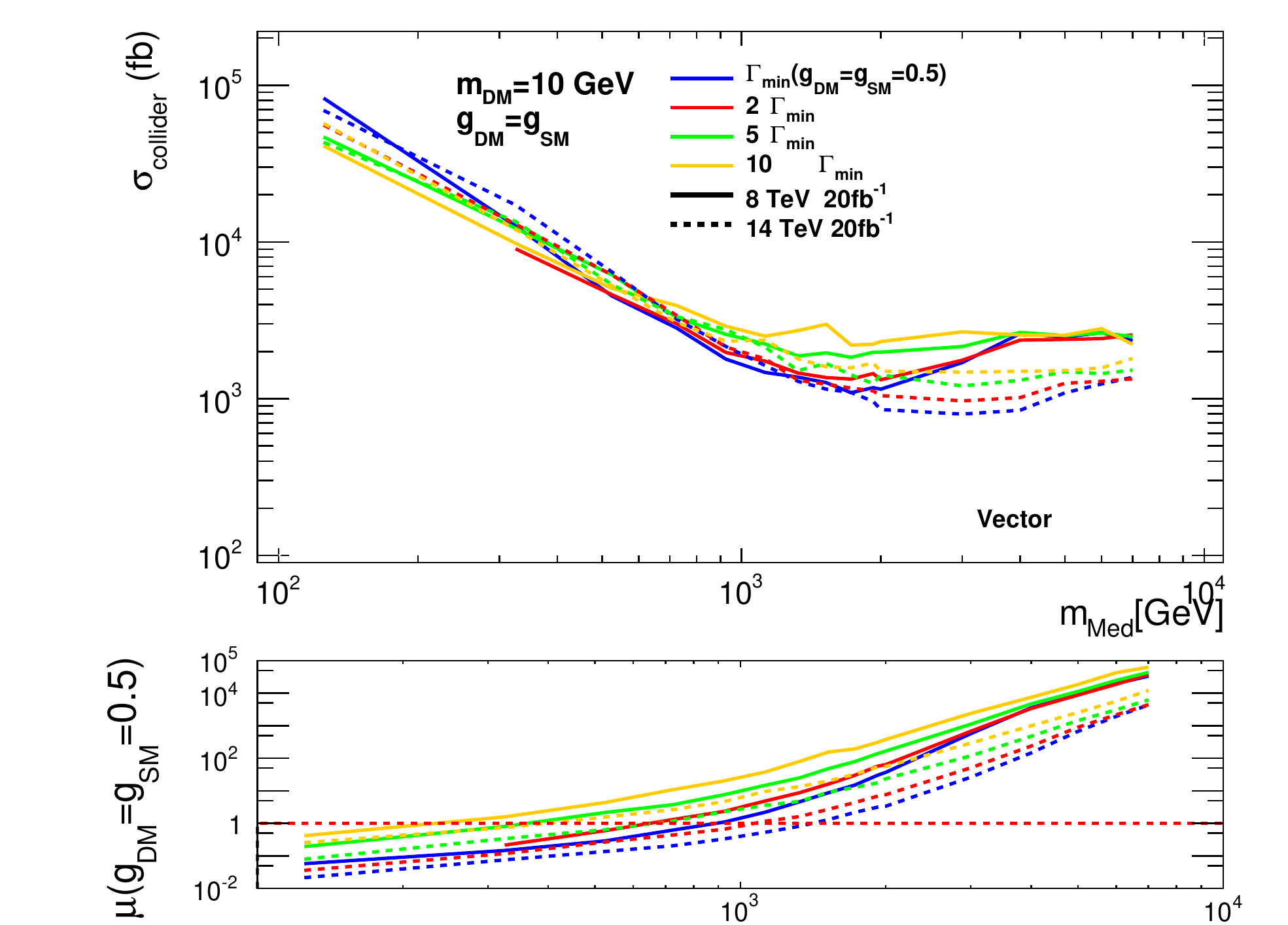}
\includegraphics[width=0.47\textwidth]{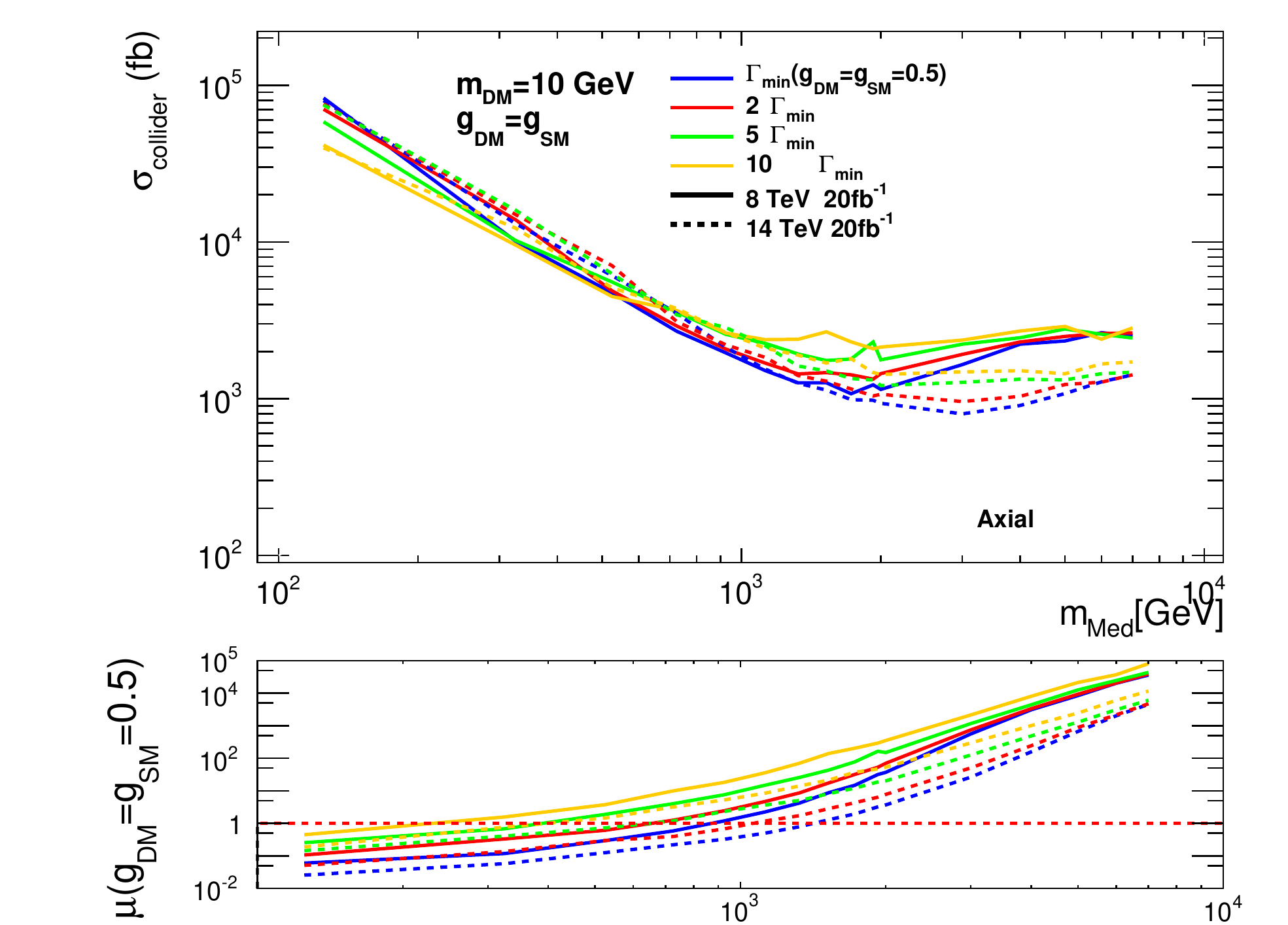}
\caption{Cross section limit bounds (projections) for LHC cross sections at 8 TeV (14) TeV for scalar (upper left), pseudo-scalar (upper right), vector (lower left) and axial vector (lower right). We chose $m_{DM} = 10$ GeV and set the coupling parameters defined in Eq.~\eqref{eq:gdef} to $g_\chi=1=g_q$ for (pseudo)-scalars and
$g_{\rm DM}=0.5=g_{\rm SM}$ for (axial)-vector mediators.
We also note that LHC exclusion bounds do not depend strongly on the
dark matter mass, $\sigma_{collider}$ refers to the MCFM cross section for a parton $p_T > 15$ GeV.}
\label{fig:LHCsigma}
\end{figure}
\end{center}

In Figure~\ref{fig:LHCsigma} we plot the limits on the LHC cross section at 8 TeV and projected limits at 14 TeV  for scalar, pseudo-scalar, vector and axial-vector mediators,
as the function of the mediator mass.  For the $\chi$-particle dark matter mass we have chosen a relatively small value of 10 GeV, although the results 
obtained for heavier dark matter were found to be similar.  The kinematics of the process is then completely specified once a coupling is set, since this fixes 
the minimal width of the mediator. For the coupling parameters we choose $g_g=g_\chi=1$ in the scalar and pseudo-scalar case, 
and $g_{SM}=g_{DM}=0.5$ for (axial)-vectors. With the kinematics of
the model fixed the properties of the model allow a limit on a cross section to be
derived. From the derived limit, a value, $\mu$, is obtained which refers to the ratio of excluded cross section with respect to the
predicted cross section as determined by the couplings and width constraint. Translating the
constraint on $\mu$ to direct constraints on the couplings, $g$,
requires propagating the cross section dependence of both the
couplings and the width. Values with $\mu<1$ typically indicate the
excluded couplings and  width are smaller than the tested
model. 

 In each plot we show the exclusion contours for different choices of the mediator width
$\Gamma_{\rm MED}$ starting with the minimal width $\Gamma_{\rm min}$ computed in each model for the given choice of parameters, 
and then scaling it upwards as $2 \times \Gamma_{\rm min}$, $5 \times
\Gamma_{\rm min}$ and $10 \times \Gamma_{\rm min}$. We quote this both
in terms of a cross section limit, and $\mu$.  For low values of the
mediator mass, we find that an exclusion of order unity in
$\mu$. Additionally, we observe that  
increasing the energy from 8 to 14 TeV results in a sizable increase in the limit for heavy ($> 1$ TeV) mediators. 

\begin{center}
\begin{figure}
\includegraphics[width=0.45\textwidth]{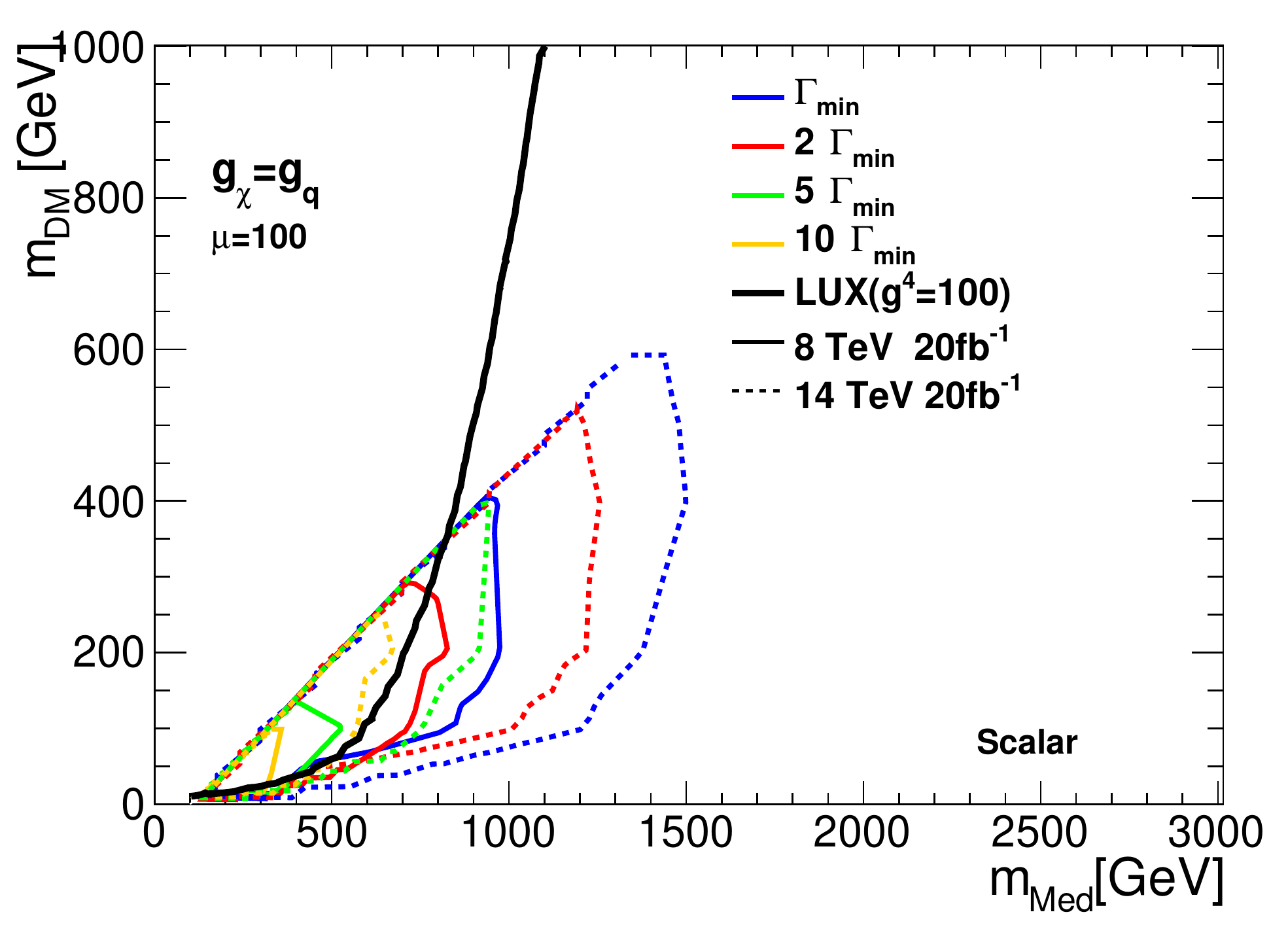}
\includegraphics[width=0.45\textwidth]{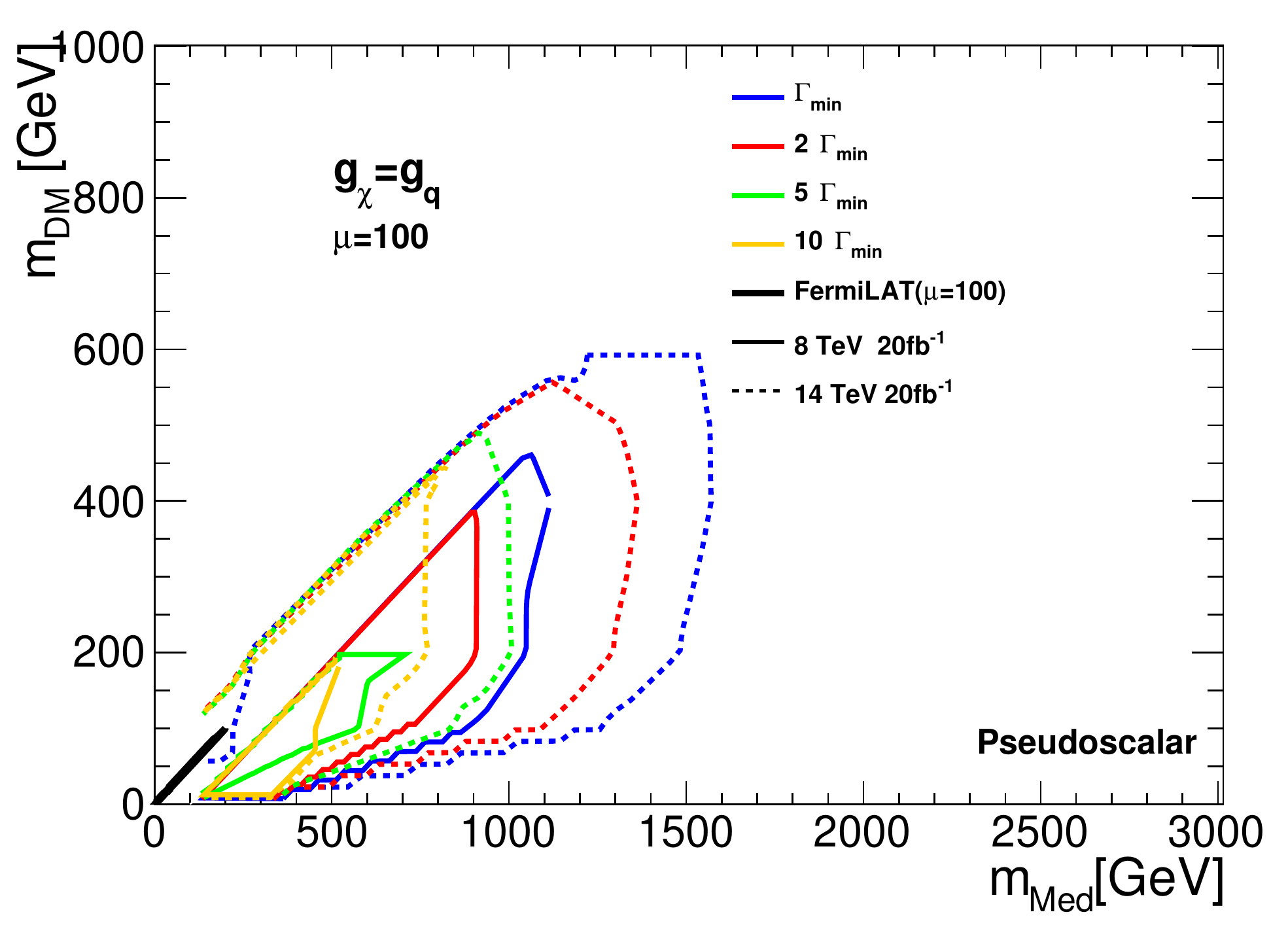}\\
\includegraphics[width=0.45\textwidth]{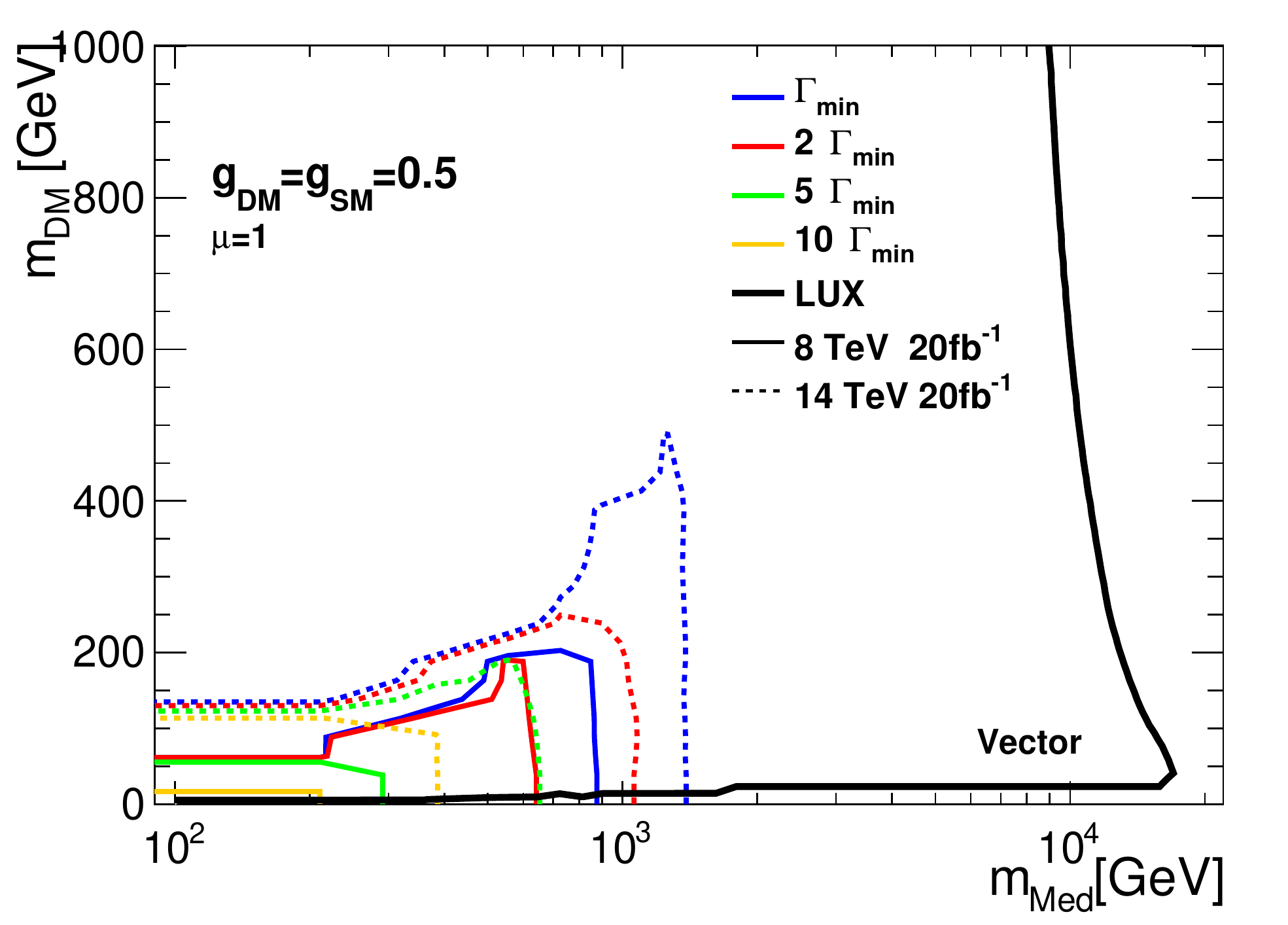}
\includegraphics[width=0.45\textwidth]{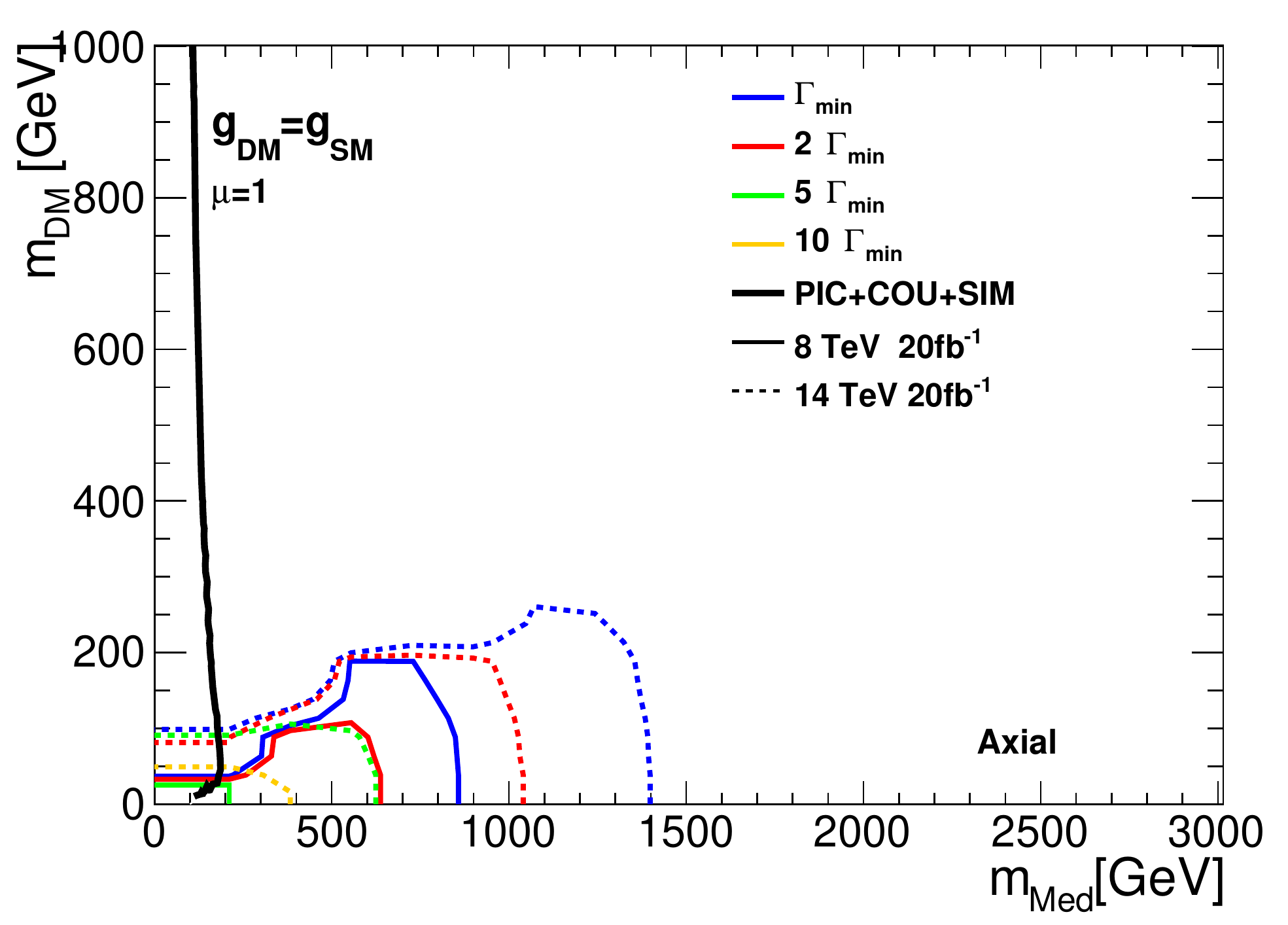}
\caption{
$m_{\rm DM}$, $m_{\rm MED}$ exclusion planes for different potential
mediators, for a variety of different mediator widths.  We show the LHC 8 and the direct detection limits, in addition 
we present expected limits at LHC 14 TeV at 20 fb$^{-1}$ as dotted lines. 
For scalar and pseudo-scalar mediators we increased the cross section by a factor of $\mu=100$ as explained in the text.
Data for direct detection results (in this and subsequent figures) comes from the LUX, \cite{Akerib:2013tjd}, PICASSO \cite{Archambault:2012pm}, COUPP \cite{Behnke:2012ys} and SIMPLE \cite{Felizardo:2011uw} experiments.
For the pseudo-scalar mediator model on the top right plot we
show the indirect detection limits using FERMI-LAT data~\cite{Ackermann:2011wa}.
}
\label{fig:DM-Med}
\end{figure}
\end{center}

In order to obtain relevant exclusion limits for our simplified models, we must compare the predicted value of the cross section for a given parameter
set against the limit set by the LHC (e.g. in Fig.~\ref{fig:LHCsigma}). We present the constrained region as a function of the dark matter and mediator mass in Fig.~\ref{fig:DM-Med}. 

To highlight 
the complementarity between the LHC limits and the direct detection (DD) experiments we
present them on the same plots in Fig.~\ref{fig:DM-Med}. 
We stress, however, that the direct comparison between the collider limits on the production of a dark sector
particle $\chi$ and the exclusion limits from direct detection and indirect detection dark matter experiments is only sensible for $\chi$ being
the cosmologically stable DM. More generally, if $\chi$ is a representative of the dark sector and is only stable on collider time scales, but not cosmologically,  DD and ID bounds are severely diluted relative to the LHC limits in Fig.~\ref{fig:DM-Med}, or even not applicable.

\begin{center}
\begin{figure}[h!]
\includegraphics[width=0.45\textwidth]{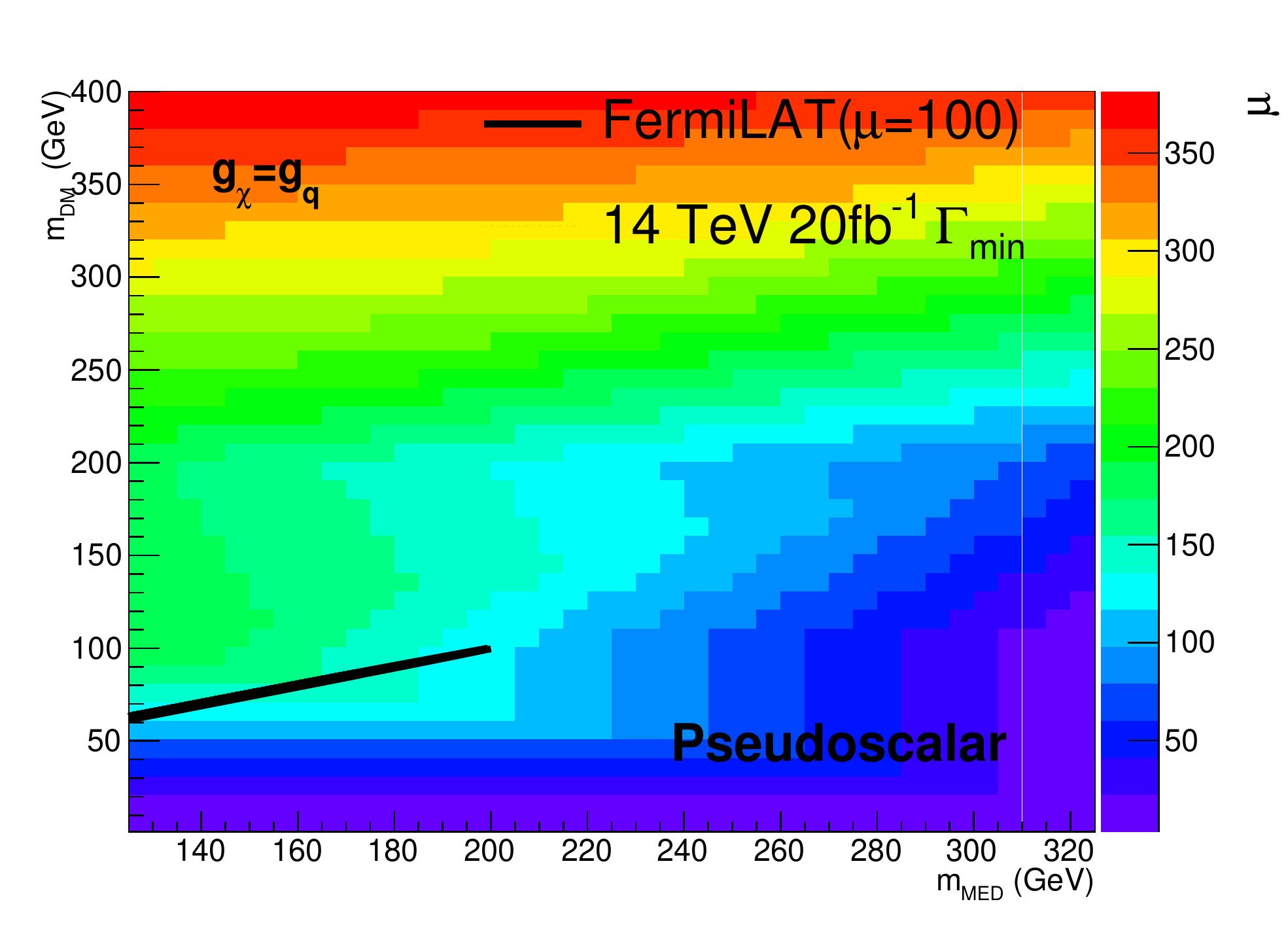} 
\includegraphics[width=0.45\textwidth]{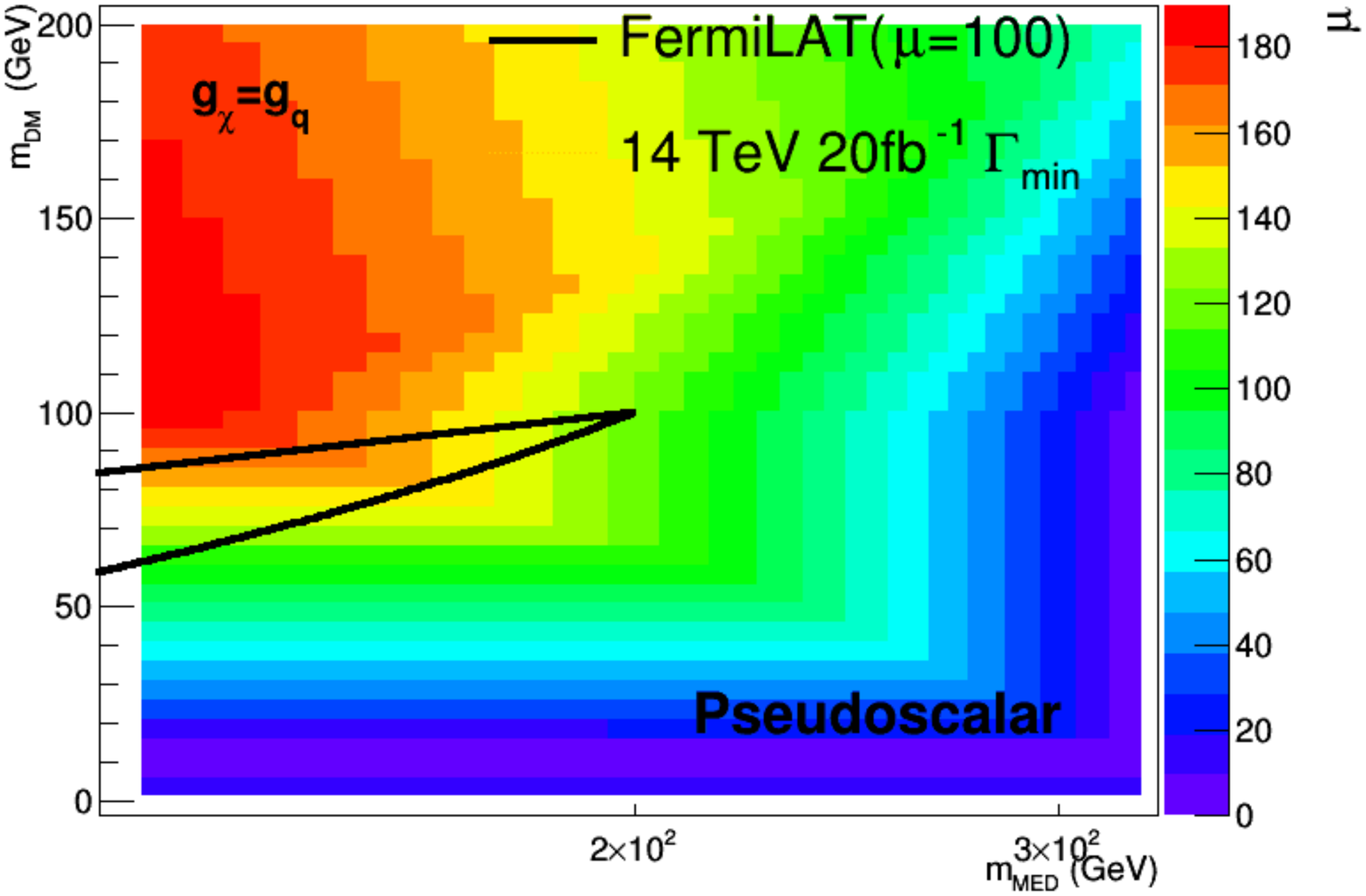}
\caption{Contour plots  illustrating the signal enhancement $\mu$-factor for pseudo-scalar mediator  to achieve a 90\% CL exclusion at 14 TeV LHC. 
The plot on the right 
shows exclusion contours for the low DM mass region, {\it cf.} top right plot in Fig.~\ref{fig:DM-Med}.}
\label{fig:zoom-mu}
\end{figure}
\end{center}

We find that for our vector and axial-vector models described above, the LHC at 8 TeV can exclude mediator masses of $<$ 1 TeV and dark matter masses of less than around 200 GeV. Our 14 TeV LHC at 20 fb$^{-1}$ projection contours improve these, roughly speaking, by a factor of two.
Note, however, that the collider limits are strongly dependent on the width of the mediator particle (due to the $\Gamma_{\rm MED}^{-1}$ scaling of the cross section), if the width is increased the limits become substantially 
weaker.\footnote{Note that due to the shape of the LHC exclusions for different limits in Fig.~\ref{fig:LHCsigma}, this is not a simple $\Gamma_{\rm MED}^{-1}$ relation.}
For direct detection experiments the limits for axial and vector are very different, which as is well established is due to the spin independent versus spin dependent 
nature of particular processes. For the spin independent vector scattering the direct detection experiments cover almost the entire available phase space in the plot (apart from 
very light DM  $< 5$ GeV scenarios which fall below the threshold energy of the detector). On the other hand, spin-dependent interactions are not as strongly constrained by 
DD experiments, here constraints are strong for light mediators, but fall off quickly for heavier mediators. The complementarity of DD and collider experiments can scarcely be made
more apparent then this plot illustrates. 

Direct and indirect detection experiments are more constrained by the mediator mass than the dark matter mass (c.f. Eqs~\eqref{eq:16}-\eqref{eq:19}), and as a result, are able 
to exclude a wider range of dark matter masses, at a cost of a smaller constraint on the mediator mass. 
For the scalar and pseudo-scalar case, no collider limit can be set with our choice of $g_{\chi}=g_{q}=1$, that is to say, that
 experiments cannot presently exclude parameters 
arising from a model in which both couplings are exactly Higgs like (for $m_{DM}=10$ GeV at least). From the lower panels of Fig.~\ref{fig:LHCsigma} it is clear that if we increase the signal by a factor of $\mu=100$ then limits can be derived. Note that this scale factor is not completely unmotivated, the width of the mediator for scalar and pseudo-scalar mediators in our model is dominated by $t\overline{t}$ decays (in particular for light DM). Therefore increasing the cross section by a factor of 100 is approximately equivalent to changing $g_{\chi}\rightarrow 10 g_{\chi}$.  Since we defined $g_{DM}=g_{\chi} m_{DM}/v$ increasing the coupling by a factor of 10 for light DM is completely plausible,
for light dark matter, e.g. $m_{DM} \lesssim 25$ GeV , such that we remain well within the perturbative regime of 
$g_{DM} \lesssim 1$,
and is basically equivalent to not requiring the mediator-$\chi$ interaction to scale like a Yukawa coupling. 
We note that the LUX and FERMI-LAT contours shown on the scalar and pseudo-scalar plots  plot of Fig.~\ref{fig:DM-Med} are obtained with the same rescaling, illustrating the challenge of searching for scalar mediators at all kinds of experiments. Extending the LHC reach to 14 TeV provides more stringent bounds, since the larger centre of mass energy allows heavier mediators to be probed. 

Plots presented in Fig.~\ref{fig:zoom-mu} illustrate the contours of the required $\mu$-factor necessary to enhance the signal for pseudo-scalar 
messenger models to set a 90\% CL at 14 TeV LHC assuming the minimal width. The question of whether a parameter point is visible at the LHC depends on the ability to separate signal processes from the background. A better background rejection will boost sensitivity independently of the signal parameterisation and the real analysis sensitivity is likely to improve by a substantial amount. Hence this figure only serves as baseline for our current sensitivity.
The plot on the right of Fig.~\ref{fig:zoom-mu} zooms into the relatively low DM mass region of the parameter space for pseudo-scalar mediators
where the data from Indirect detection experiments become relevant. 

\begin{center}
\begin{figure}[b]
\includegraphics[width=0.45\textwidth]{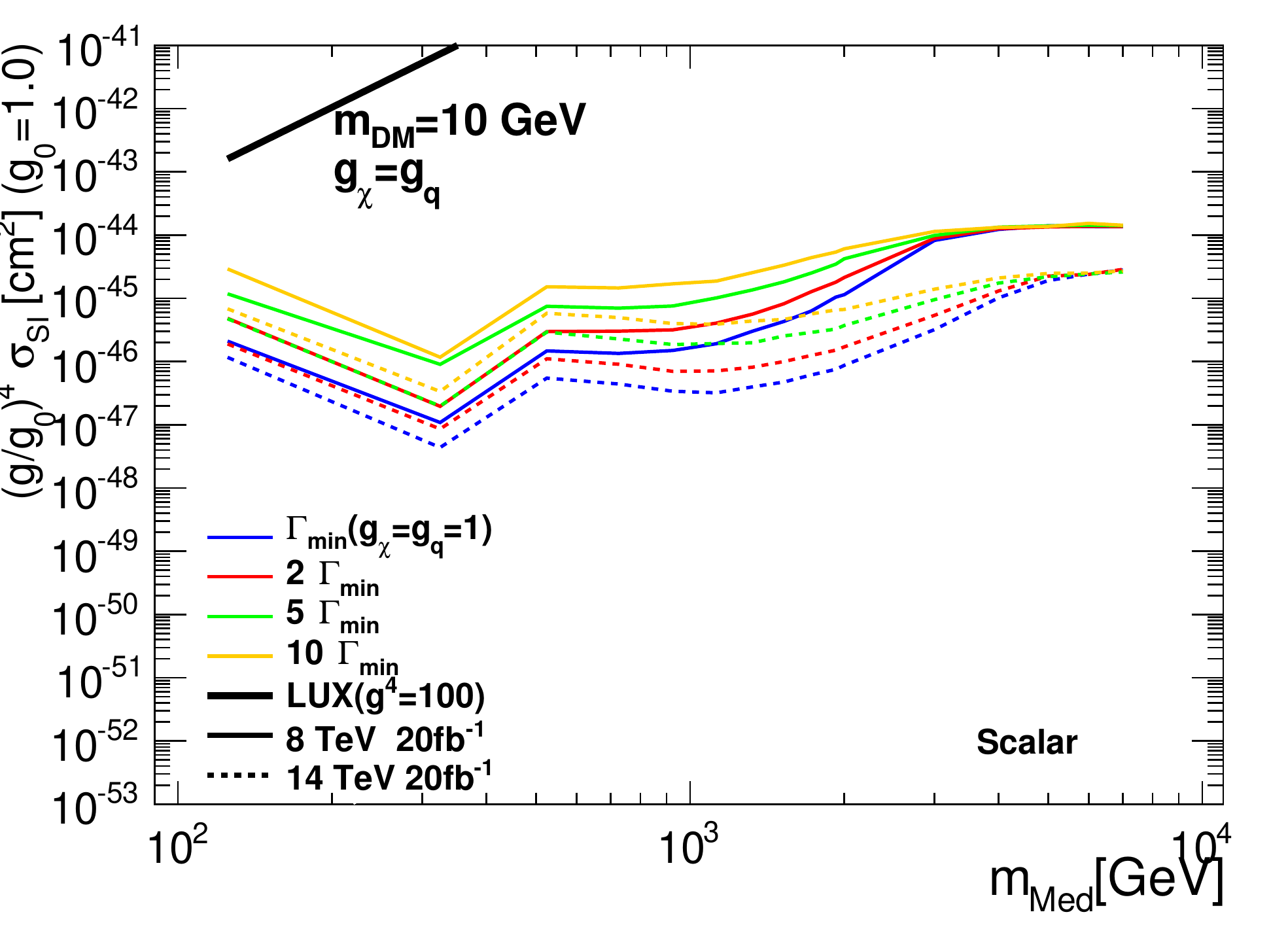}
\includegraphics[width=0.45\textwidth]{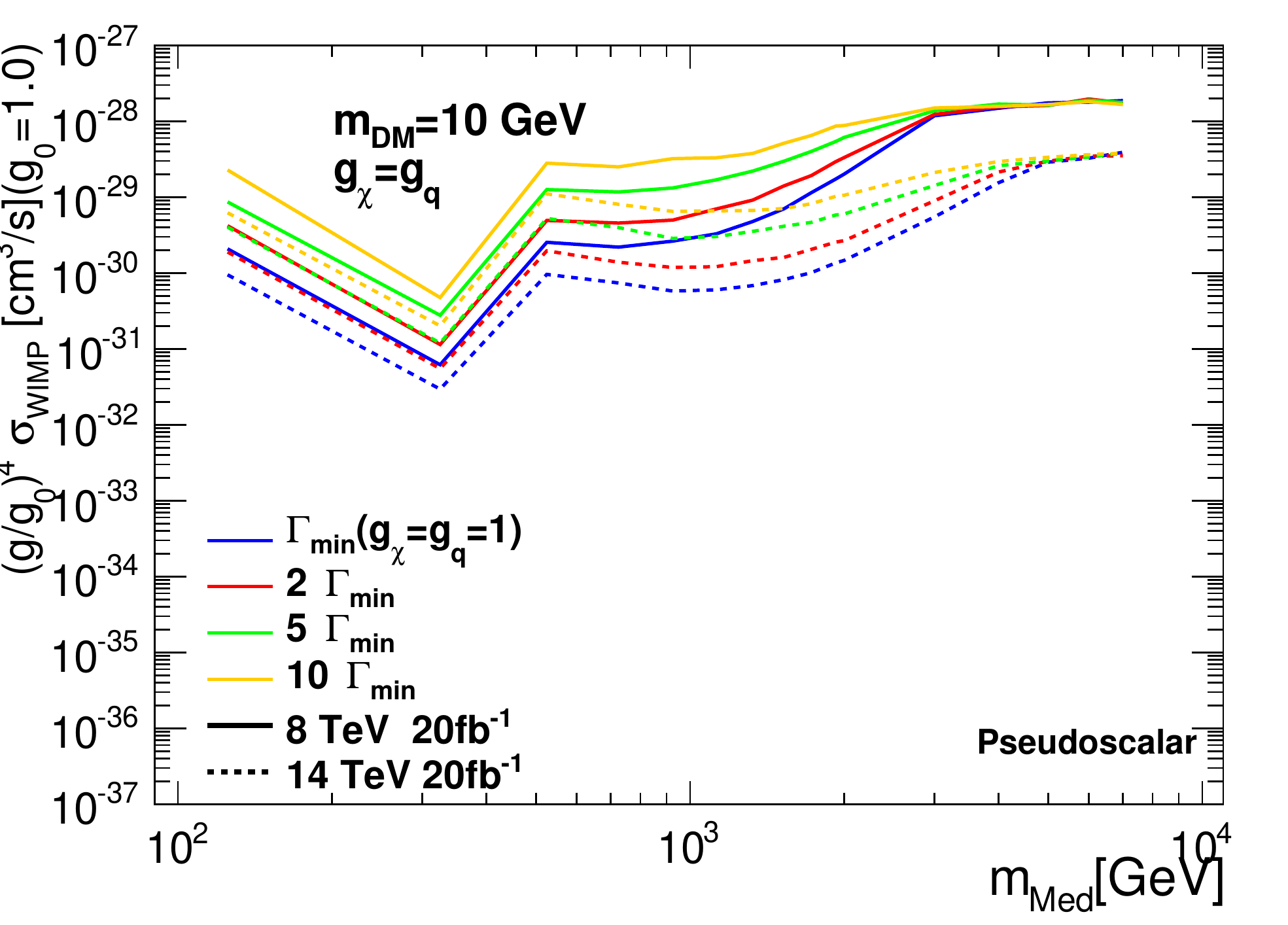}
\includegraphics[width=0.45\textwidth]{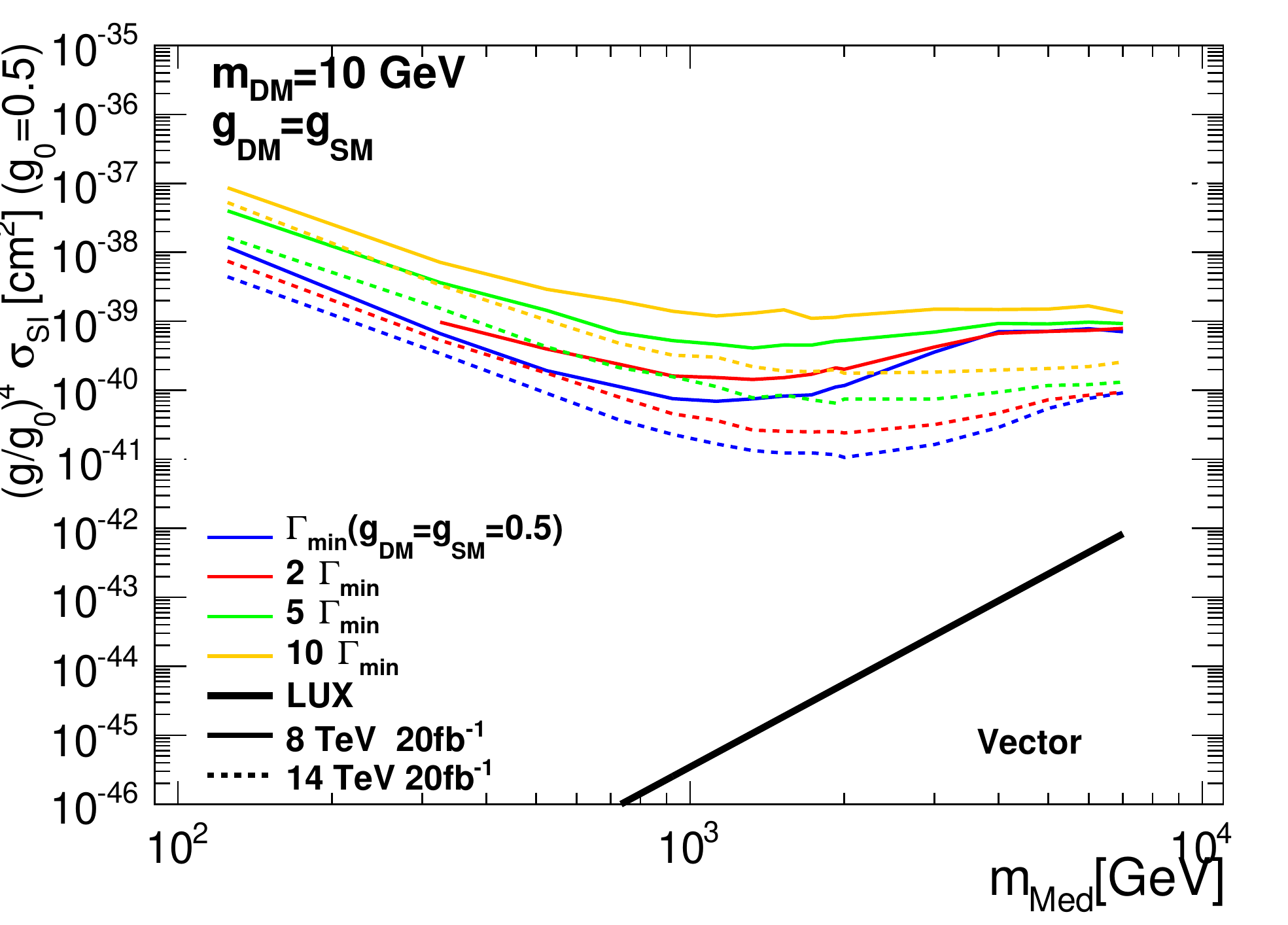}
\includegraphics[width=0.45\textwidth]{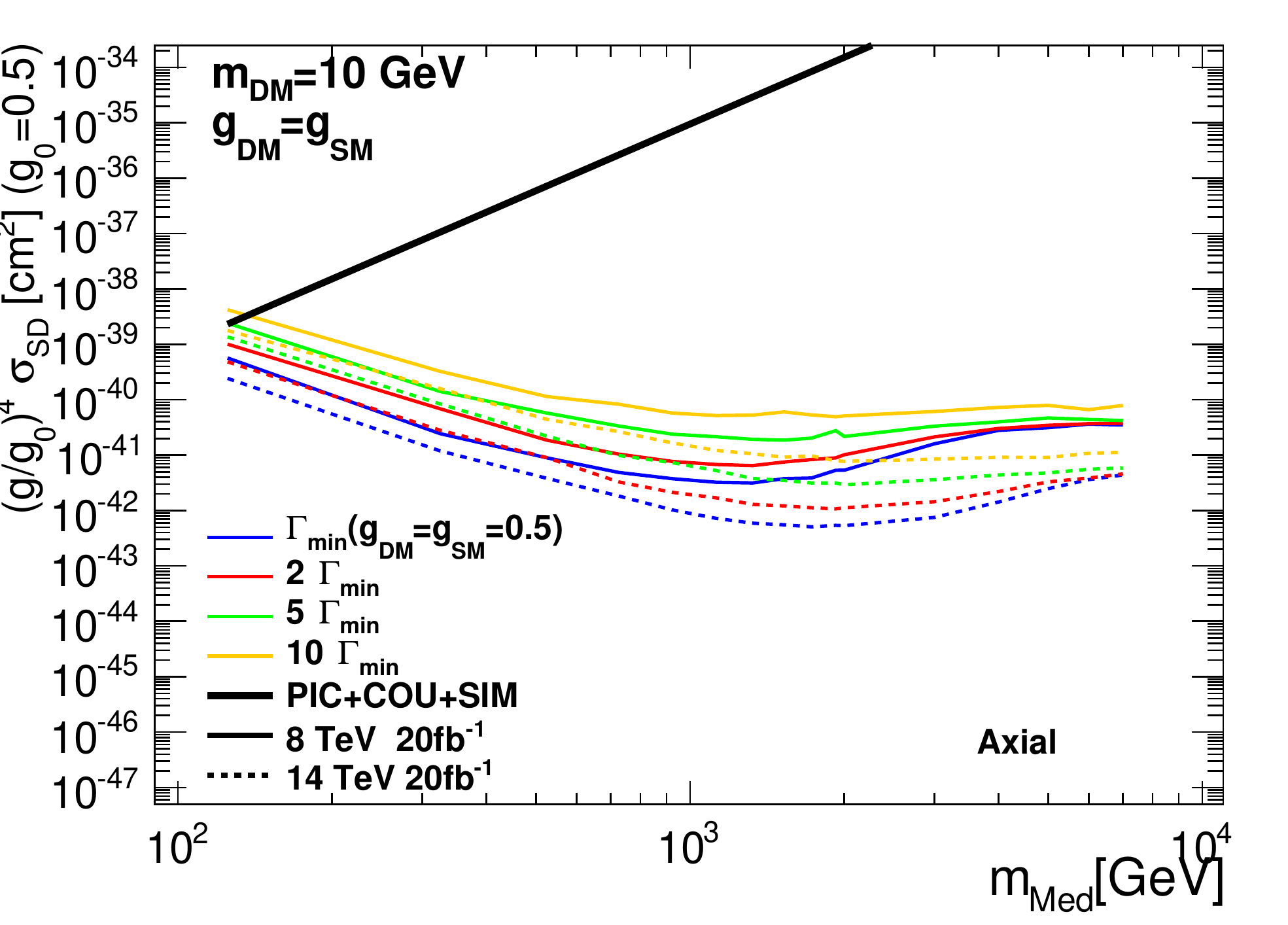}
\caption{Cross sections limits (projections) in terms of those reported by DD experiments, we show exclusions in terms of the spin-dependent and spin-independent 
cross sections against mediator mass for scalar, vector and axial-vector mediators. For the pseudo-scalar mediator we plot the DM pair annihilation cross-section.
The corresponding limit from Indirect Detection experiments is too weak to show for the pseudo-scalar messenger.}
\label{fig:DD-ID}
\end{figure}
\end{center}

\begin{center}
\begin{figure}
\includegraphics[width=0.45\textwidth]{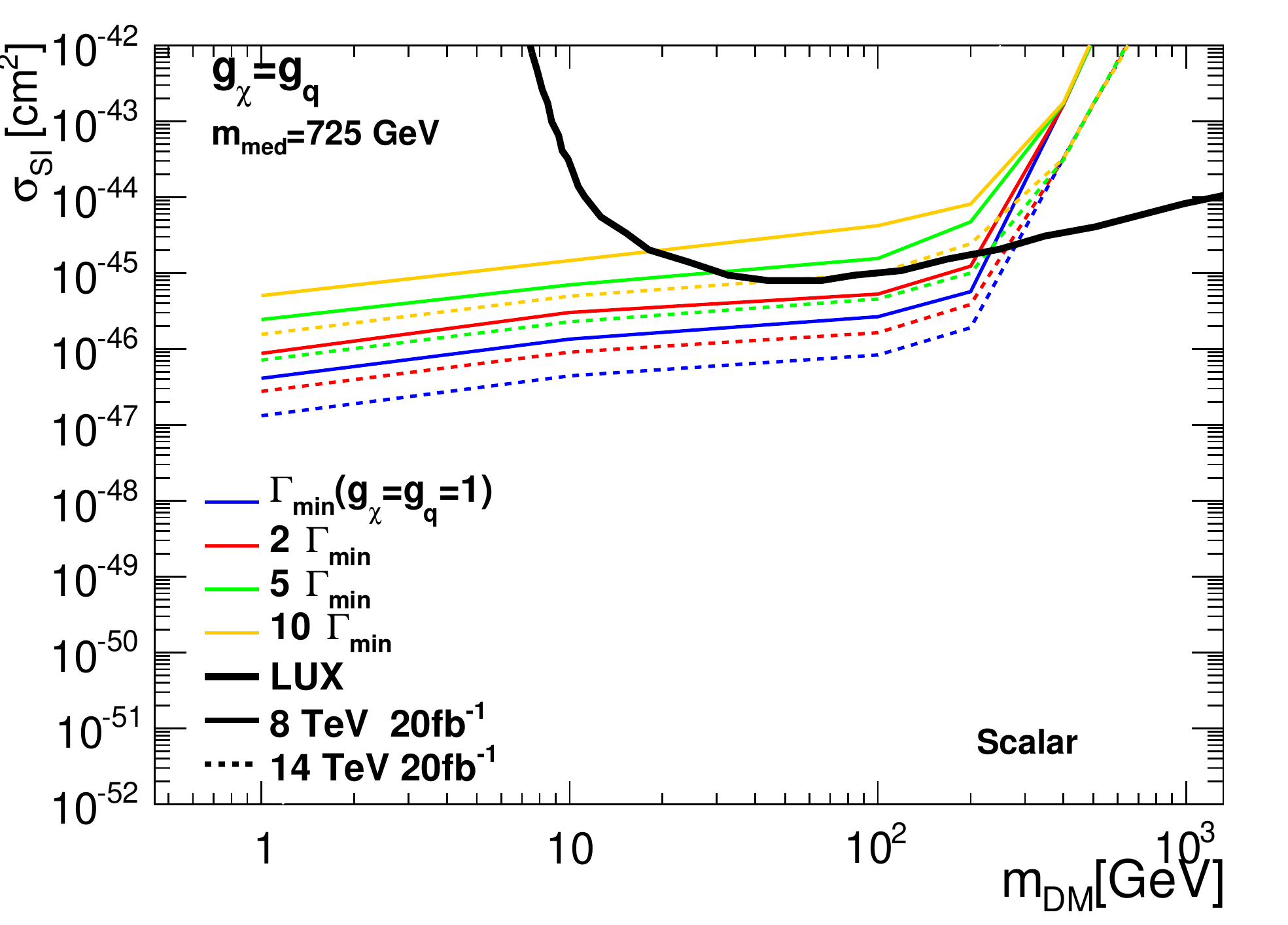}
\includegraphics[width=0.45\textwidth]{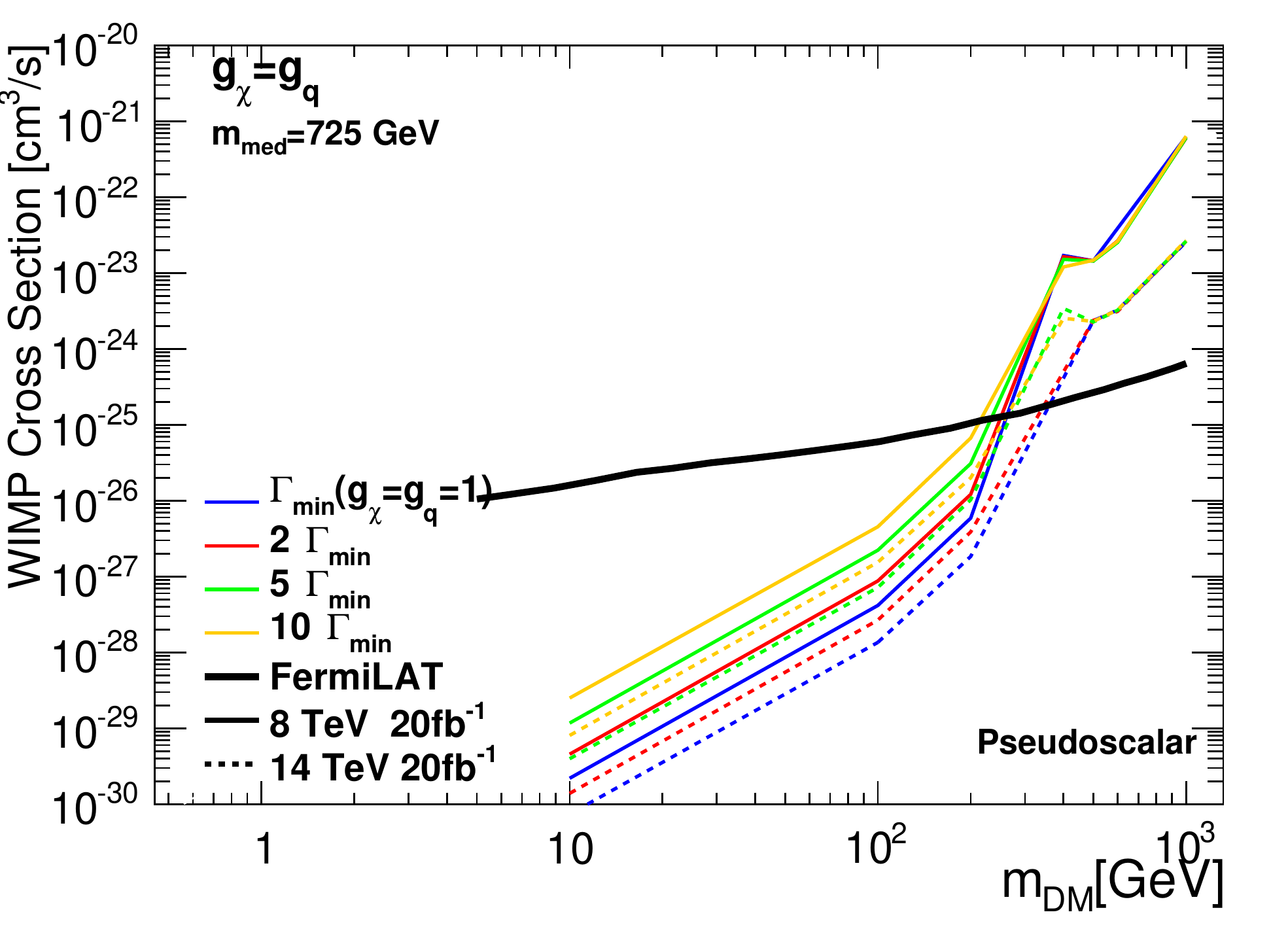}
\includegraphics[width=0.45\textwidth]{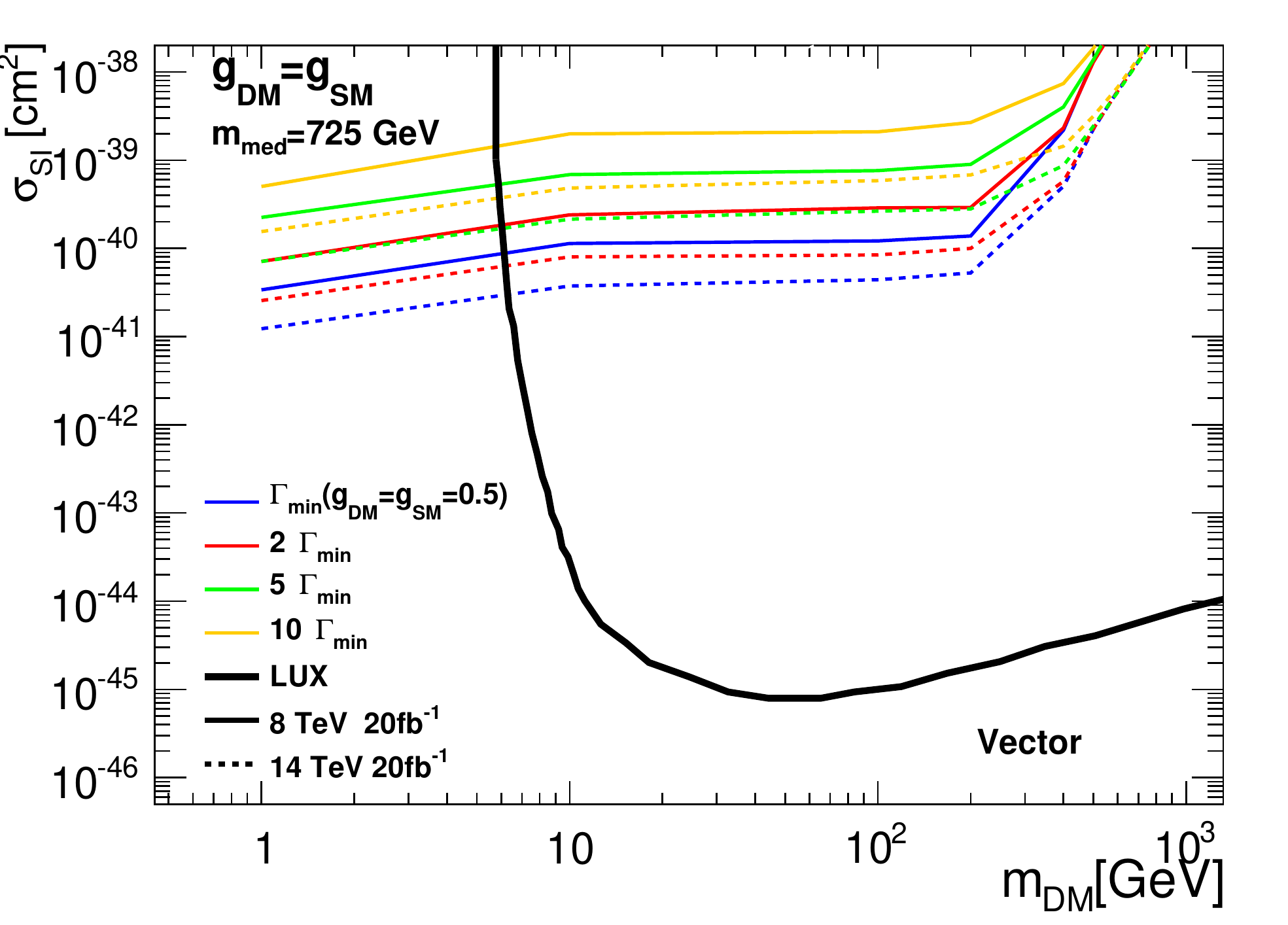}
\includegraphics[width=0.45\textwidth]{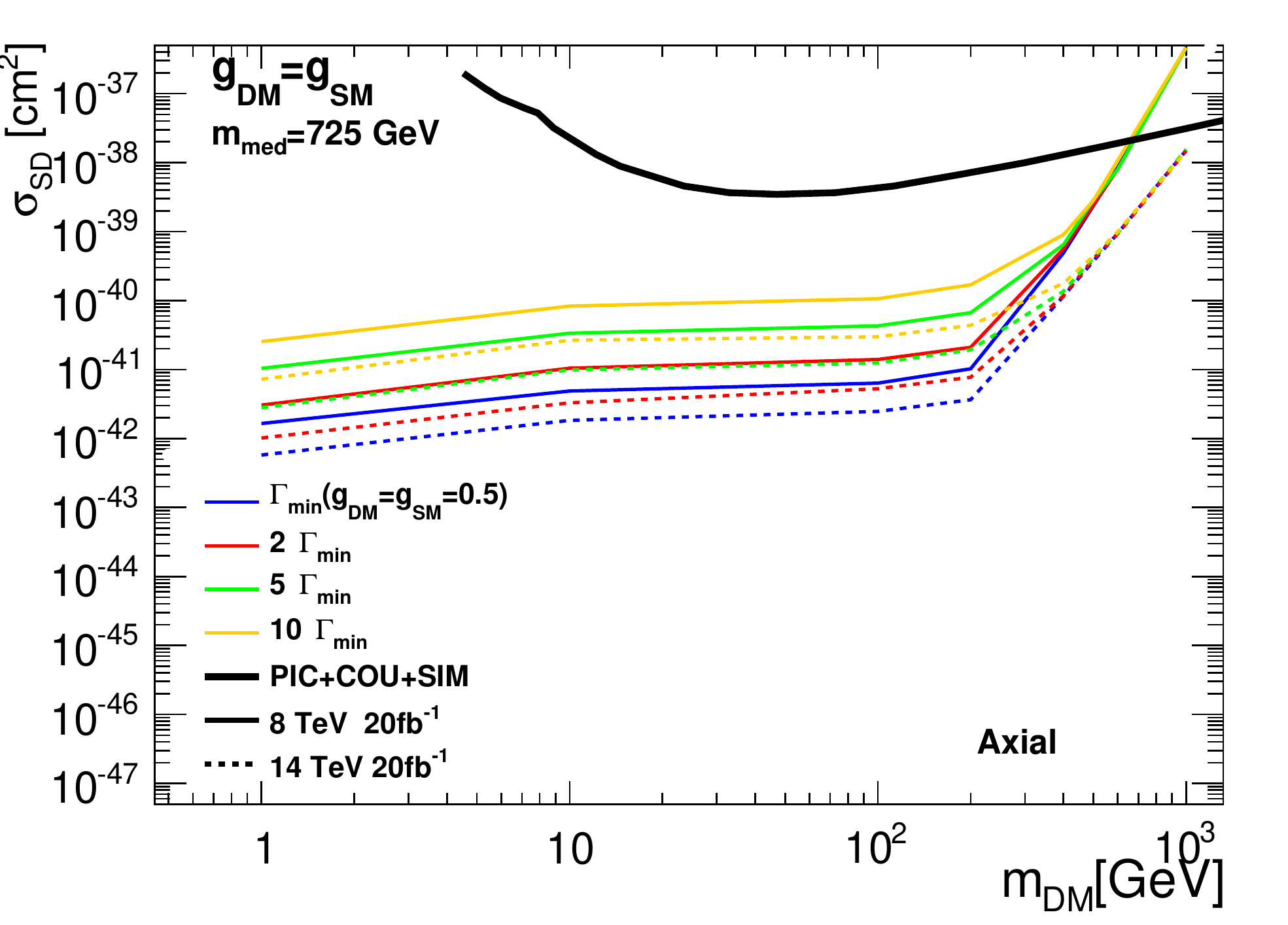}
\caption{ Exclusion contours for the spin-dependent and spin-independent cross sections 
as in Fig.~\ref{fig:DD-ID}, now plotted as functions of the dark matter mass. For the pseudo-scalar mediator model we
show the indirect detection limits (using FERMI-LATdata~\cite{Ackermann:2011wa}). For the pseudo-scalar we show $95\%$ C.L. exclusion limits, while we show limits at $90\%$ C.L. for the other mediators.}
\label{fig:DD-ID-med}
\end{figure}
\end{center}

Finally in Figs.~\ref{fig:DD-ID} and ~\ref{fig:DD-ID-med}
we show the plots in terms of the spin-dependent and the spin-independent DM--neutron cross sections 
for a more traditional comparison of collider limits in terms of our simplified models with the limits/projections  from 
the direct and indirect detection experiments computed using cross section formulae in Sec.~\ref{sec:DDandID}. We compare the results in the $\sigma-m_{MED}$ and $\sigma-m_{DM}$ planes. The general pattern of Fig.~\ref{fig:DM-Med} is reproduced, with spin-independent results from LUX providing the strongest bounds, for the axial and 
scalar cases the example we have chosen to illustrate is for a mediator which is too heavy to be efficiently probed at DD experiments, resulting in stronger bounds from the LHC. 

\begin{center}
\begin{figure}
\includegraphics[width=0.45\textwidth]{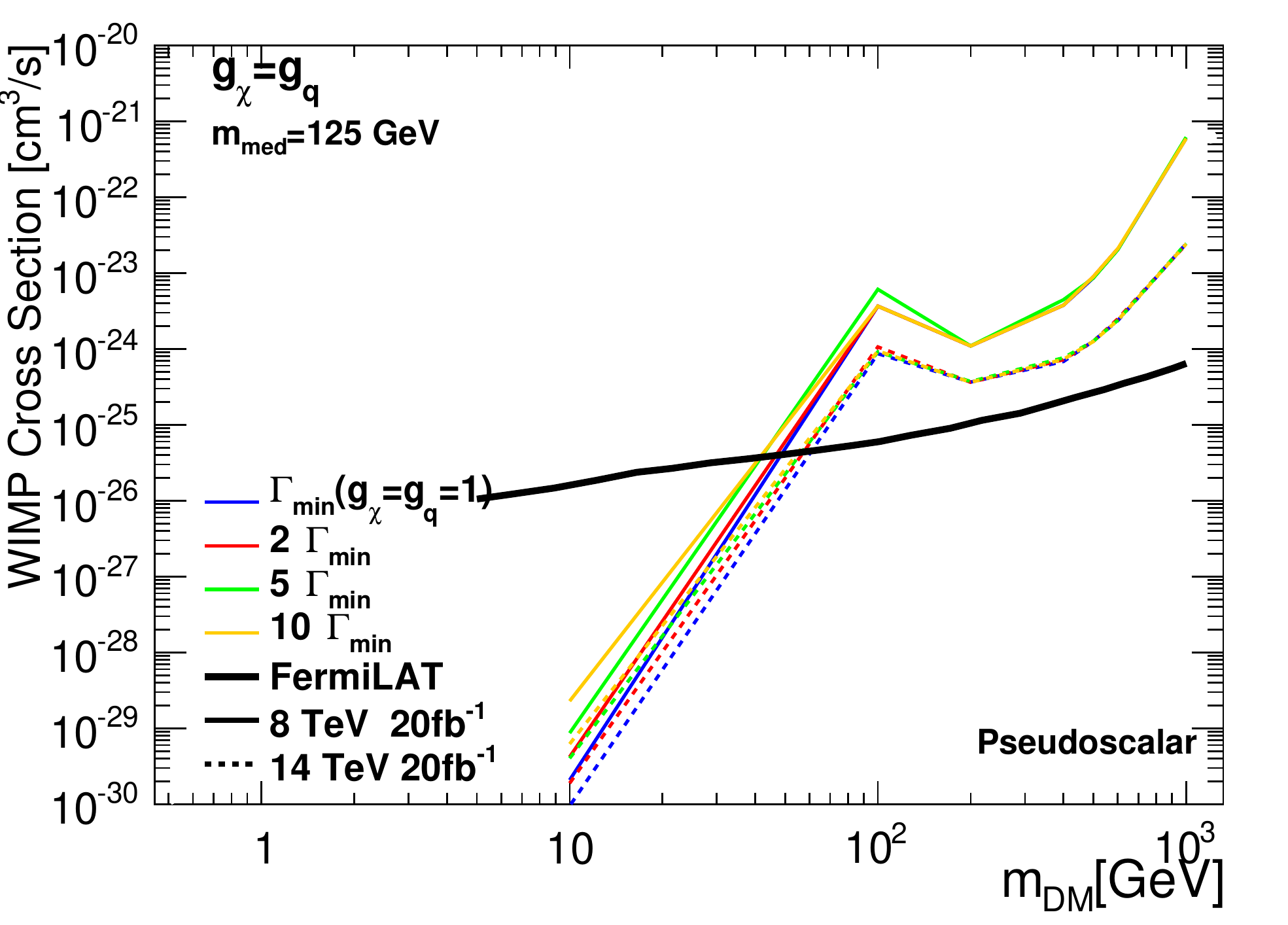}
\caption{ The spin-independent cross section
as in top right Fig.~\ref{fig:DD-ID-med} for 125 GeV pseudo-scalar mediator mass.}
\label{fig:ID-med}
\end{figure}
\end{center}

In Fig.~\ref{fig:ID-med} we show the spin-dependent cross-section limits deduced from the LHC
projections and FERMI-LAT now for a light 125 GeV mediator.

\section{Effects of heavy new physics on mediator production}
\label{sec:nploop}

In this section we investigate potential BSM effects which may alter the production of the dark sector mediator.
In particular we focus on additional heavy degrees of freedom, which are charged under $SU(N_c)$. 
Since we assume that these new degrees of freedom are heavy we can work in the limit in which 
the new states are integrated out (however, we stress that the mediator remains a propagating particle). 
This is achieved by including the following interaction in our simplified model Lagrangian in Eq.~\eqref{eq:LS}, 
\begin{eqnarray}
\mathcal{L}_{\mathrm{EFT}} = g_{g}\frac{\alpha_s}{12\pi v} S \; {\rm{Tr}} \,({G^{\mu\nu}G_{\mu\nu}})
\label{eq:eftLS}
\end{eqnarray}
For simplicity we have focused on the scalar mediator, and parameterized the Lagrangian in terms of a rescaled Higgs 
EFT dimension-5 operator (in which the rescaling factor is $g_g$). Our extended simplified model  now has an additional 
parameter $g_g$, resulting in a total of 6 free parameters. 

In order to make predictions for the resulting model we need to extend the existing implementation of this process in MCFM~\cite{Fox:2012ru},
which is based upon modified matrix elements for Higgs production (computed originally in ref.~\cite{Ellis:1987xu}). The inclusion 
of Eq.~\eqref{eq:eftLS} in the Lagrangian results in a new term which interferes with the top loop contribution at the amplitude level. Accordingly we have recomputed the production amplitude $gg\rightarrow g+S$ and $q\overline{q}\rightarrow g+S$ in terms of helicity amplitudes. The results for these amplitudes, which to the best of our knowledge, have not been reported elsewhere, are included in the Appendix. Representative Feynman diagrams from this extended model are the first two diagrams in Fig.~\ref{fig:feyn}, with the first
representing the new BSM contribution assumed to be induced by heavy colored particles.

\begin{center}
\begin{figure}
\includegraphics[width=0.35\textwidth]{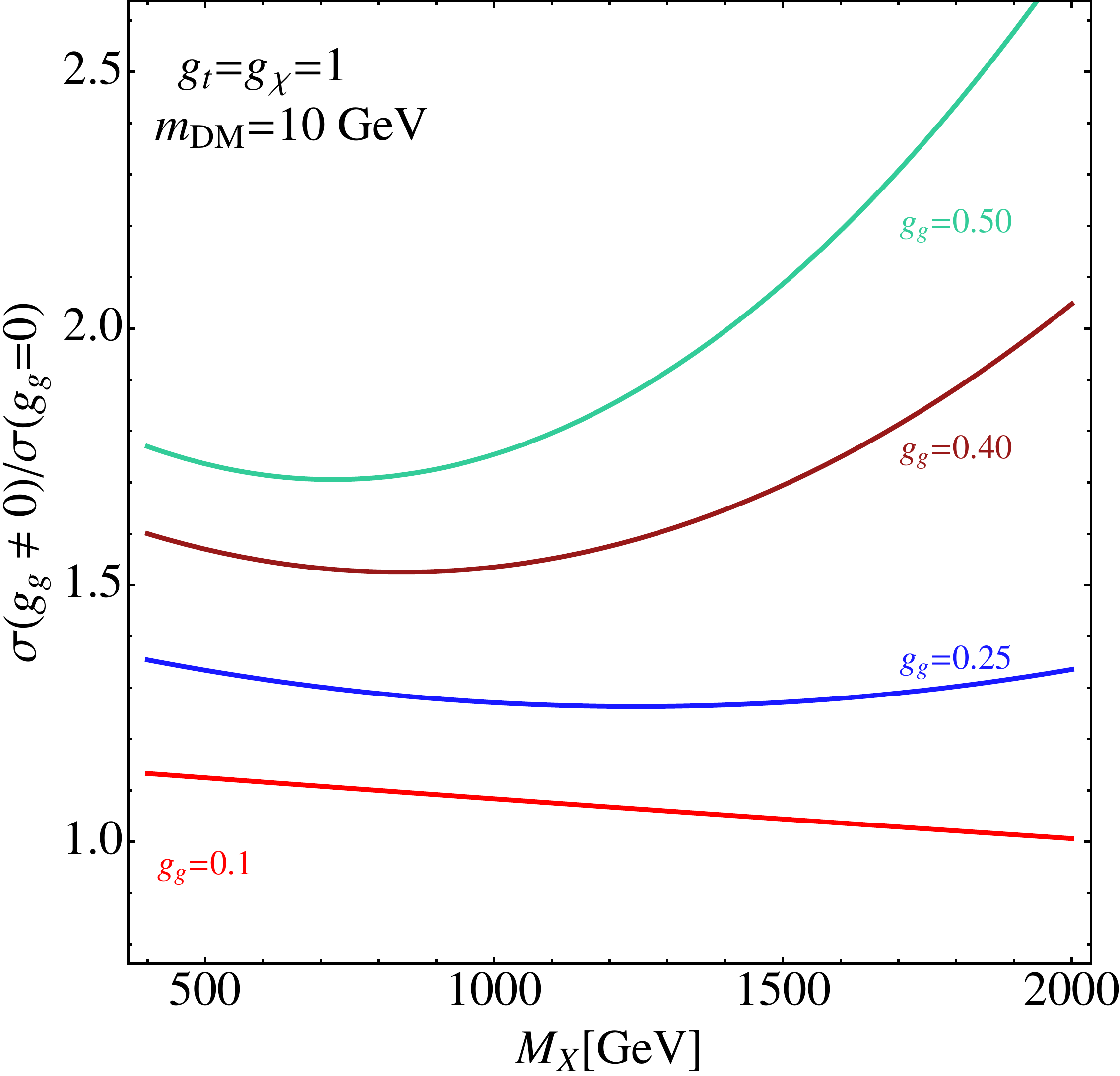}
\includegraphics[width=0.35\textwidth]{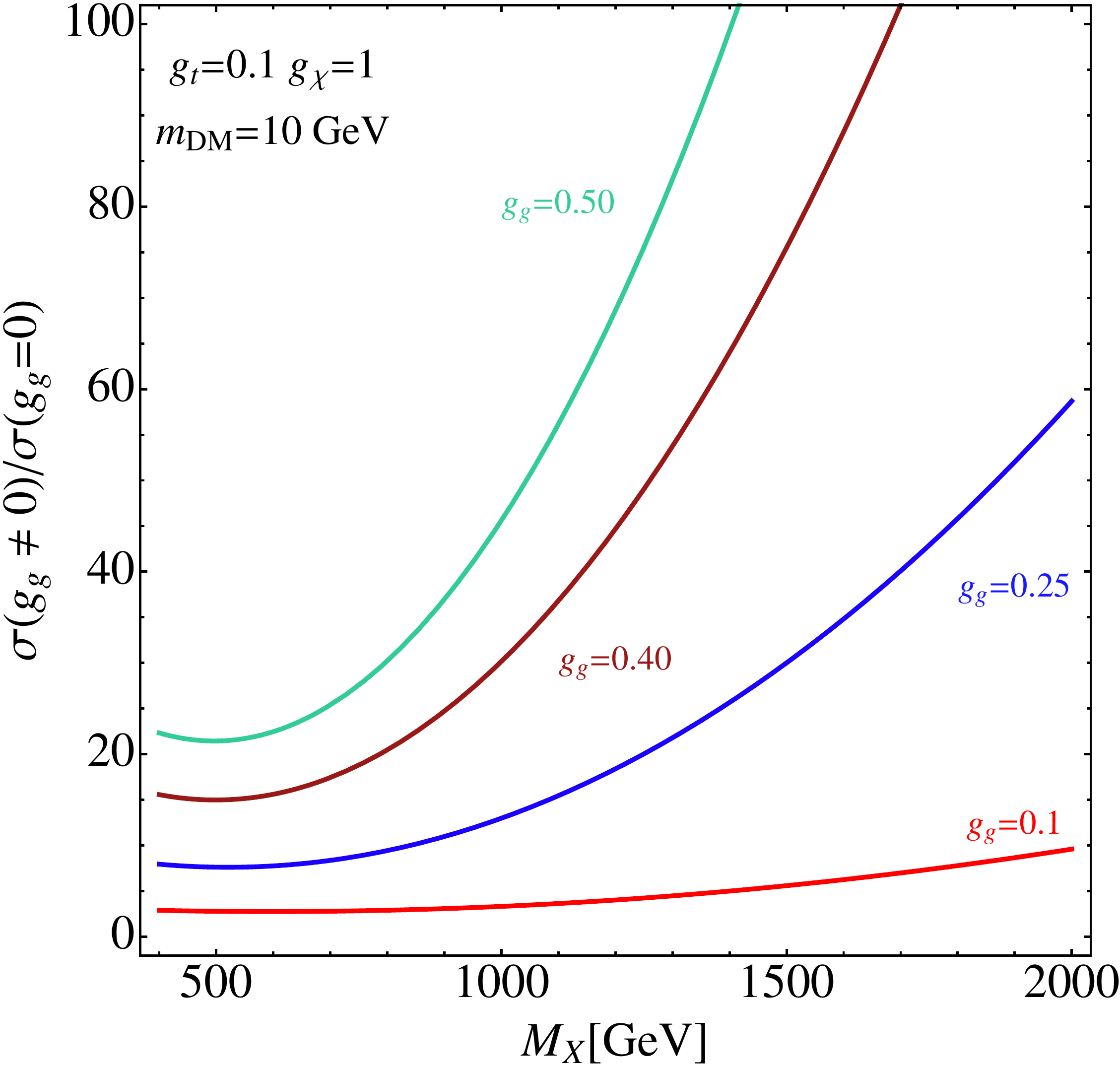}
\caption{Cross sections ratios describing the impact of the 5-dimensional contact interaction between gluons and the scalar mediator. The ratio is computed for fixed $g_\chi=1$, the plot on the left also sets $g_t=1$, whilst on the right the top-mediator coupling is weakened to $g_t=0.1$. In both instances the width is evaluated as the minimal width. The dark matter mass is fixed at $m_{DM}=100$ GeV, the mediator mass is varied. }
\label{fig:xsratioggeft}
\end{figure}
\end{center}

\begin{center}
\begin{figure}
\includegraphics[width=0.4\textwidth]{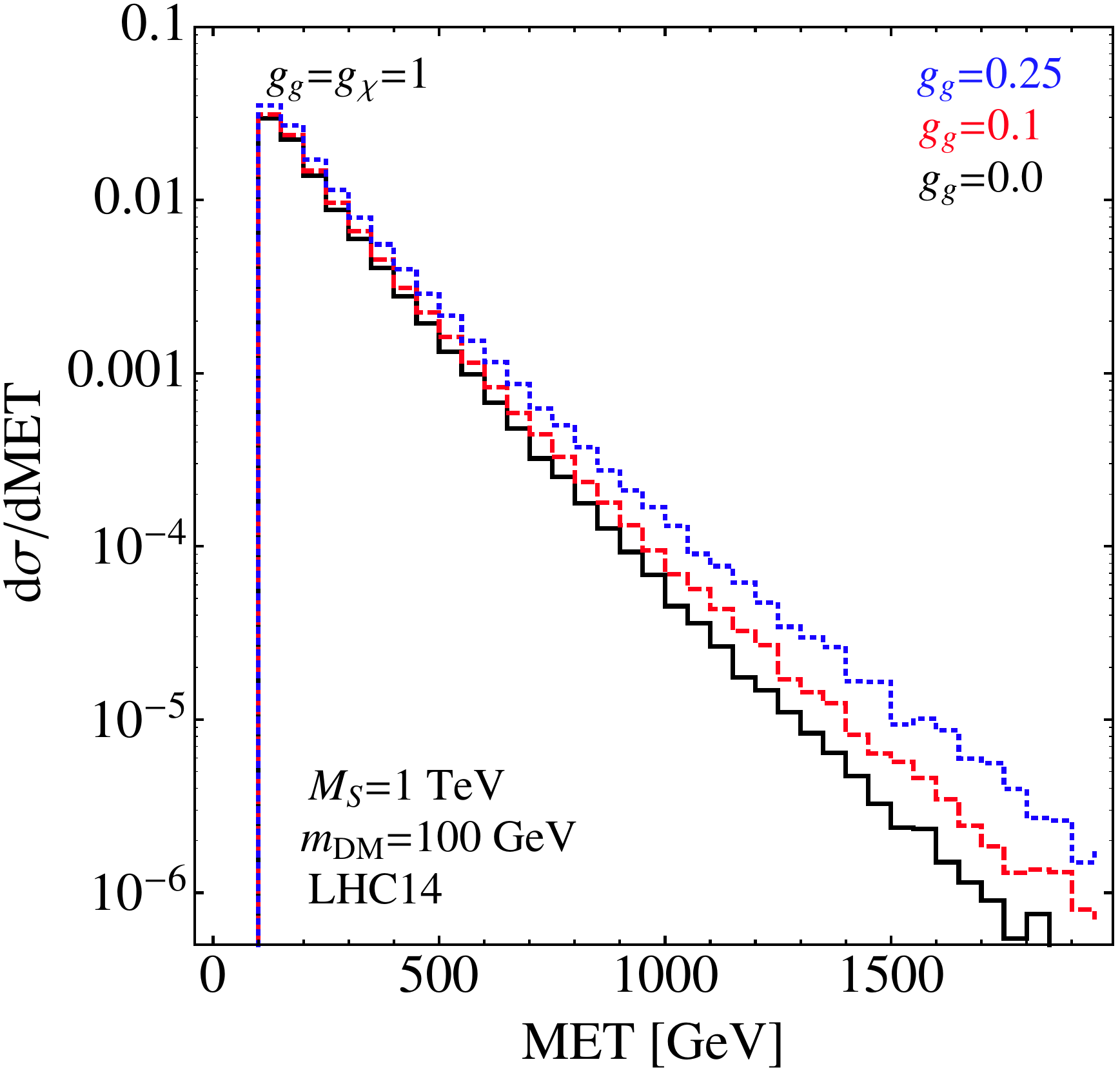}
\includegraphics[width=0.4\textwidth]{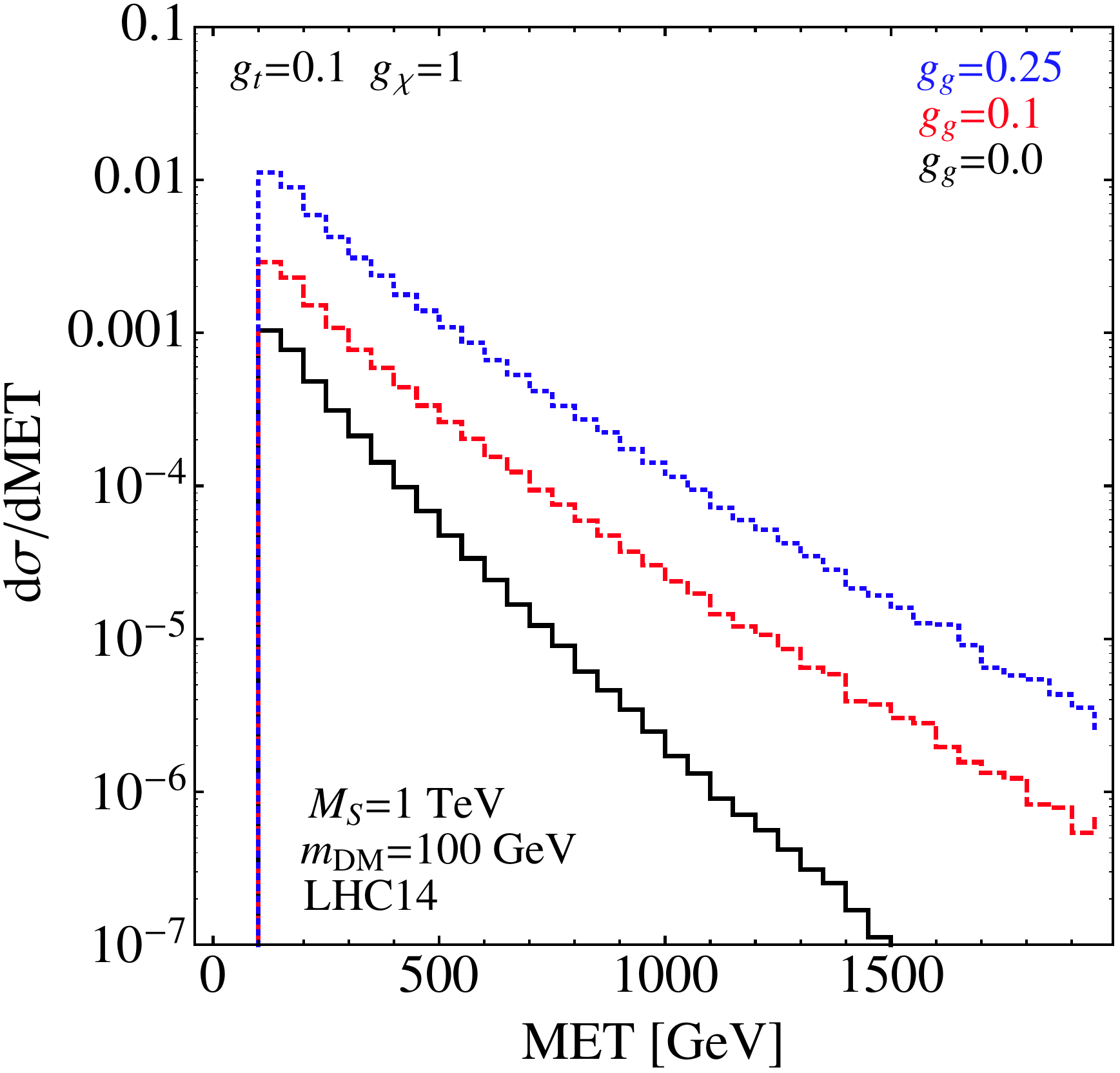}
\caption{The missing transverse energy (MET) differential distribution showing parton-level predictions obtained using several choices of $g_g$ for a benchmark scenario in which the dark matter particles 100 GeV and the mediator is 1 TeV. The plot on the left corresponds to the democratic choice of mediator-top and mediator-DM coupling ($g_t=g_\chi=1$, whilst that on the right corresponds to the case in which $g_t=g_\chi/10=0.1$.}
\label{fig:histoMETgg}
\end{figure}
\end{center}

In Fig.~\ref{fig:xsratioggeft} we present the cross section ratio $\sigma (g_g \ne 0)/\sigma(g_g=0)$ computed in the extended simplified model. 
Cross sections are obtained using the basic cuts described in section~\ref{sec:searches}, i.e. we require $p_{T,j_1} \geq 110$ GeV and $|\eta| < 2.4$. Since these are LO cross sections, an equivalent cut of 110 GeV is imposed upon the MET. Cross sections are presented for the 14 TeV LHC, and CTEQ6L1~\cite{Lai:2010vv} PDF sets have been used, the renormalization and factorization scale are set to $\mu=m_{\chi\overline{\chi}}$.
Parametrically we expect a prediction which is of the following form, $ a_0+a_1 g_g + a_2 g_g^2$, and the curves presented in Fig.~\ref{fig:xsratioggeft} illustrate this functionality. We have chosen to illustrate two benchmark points, which correspond to a situation in which the (Yukawa adjusted) mediator-top and mediator-DM coupling are set equal ($g_t=g_{\chi}=1$), and a second scenario in which the coupling to the top is damped $g_t=g_{\chi}/10=0.1$. Clearly in the second scenario we expect a significantly bigger 
contribution from the heavy particles in the loop, since the top is suppressed. This is illustrated in Fig.~\ref{fig:xsratioggeft}, for which we see enhancements of order 1-3 for the democratic model, and 10-100 for the top suppressed model.  The second case is analogous to the situation in the production of a SM Higgs, in which the propagating  $b$ quark loops are suppressed relative to the top quark contribution, which can be evaluated in the heavy top limit. We note that the shapes of the curves in Fig.~\ref{fig:xsratioggeft} are dominantly due to the relative importance of $g_t$ compared to $g_g$, and the mediator mass. The dependence on $g_\chi$ cancels in the ratio, as does the decay matrix element to DM, therefore, although we chose a benchmark mass of $10$ GeV, Fig.~\ref{fig:xsratioggeft} would not be changed if a  different $m_{DM}$ were used. 
 
Under the assumption that the scalar mediator proceeds through a portal interaction with the Higgs boson, $g_g$ plays the role of a mixing angle, if $g_g$ becomes large, significant deviations from the SM Higgs should be expected, and have not been observed. If instead the scalar is allowed to couple to the heavy degrees of freedom in an arbitrary way then we interpret $g_g$ as a rescaling factor which relates the coupling and mass scale of the dark sector, to that of the EW scale, i.e.
\begin{eqnarray}
g_g \sim \frac{g_{NP}}{\Lambda_{NP}} \frac{v}{g_w}
\end{eqnarray}
In order for the EFT to be valid we need to ensure that the kinematic distributions are probed at scales less than $\Lambda_{NP}$, we present the differential distribution for the missing transverse momentum in Fig.~\ref{fig:histoMETgg}. At 14 TeV the tail of the distribution probes scales of around a few TeV, setting $\Lambda_{NP}= 2$ TeV and assuming $g_{NP}$ is $\mathcal{O}(1)$ we see that the maximum $g_g$ which can be safely probed at the 14 TeV LHC is around $g_g < 0.3$. 
Figure~\ref{fig:histoMETgg} illustrates the usual result that higher dimensional operators are relatively less suppressed at high energies compared to their four dimensional counterparts, as a result the impact of the $g_g$ pieces can be reinterpreted as momentum dependent form factor which modifies the four-dimensional Lagrangian.  

The results presented in this section suggest that, should a propagating resonance be found in the mono-jet channel at the LHC, coupling constraints on loop-induced heavy sector particles can be investigated, of which values approximately $g_g < 0.3$ correspond to theories in which the EFT prescription is viable. These constraints may shed light on extended sectors in the BSM theory which contain heavy colored particles. In addition, if Run II searches based on the simplified models defined in Sec.~\ref{sec:model} lead to null results, then one can also test models in which the scalar mediator, and putative dark matter particles are light, but only couple to the SM through a heavy colored messenger. These instances correspond exactly to the situation in which $g_g$ is non-zero, but $g_{t} \ll g_{g}, g_{\chi} $.  In theories with this coupling structure the EFT becomes the dominant production model, and although at the cost of an additional parameter, $g_g$ should be included in the simplified model.  

\section{Conclusions}
\label{sec:conclusion}

We have defined benchmark or simplified models for dark particle searches for the cases of scalar, pseudo-scalar, vector and axial-vector
mediators between the SM and dark sectors. These models are defined by the interaction in Eqs.~\eqref{eq:LS}-\eqref{eq:LA} and 
\eqref{eq:eftLS}. Apart from the choice of mediator type these models are characterised in our approach 
by the following free parameters:
\begin{enumerate}
\item mediator mass $m_{\rm MED}$
\item mediator width $\Gamma_{\rm MED}$ 
\item dark matter mass $m_{\rm DM}$
\item effective coupling parameter $g_q  \cdot g_\chi$ for scalar and pseudo-scalar mediators \eqref{eq:gdef}; 
and $g_{\rm SM} \cdot g_{\rm DM}$ for axial-vector and vector mediators. 
\end{enumerate}
In our examples here we chose to study democratic scenarios in which the couplings in the dark sector and SM were equal, although this 
need not be the case this reduces the degrees of freedom from 5 to 4.
We have implemented simplified  models based on these parameters into a fully flexible (and public) Monte Carlo code, MCFM. We used MCFM 
to generate signal events, which were processed through event and detector simulation for the 8 and 14 TeV LHC. We were then able to 
use a recent CMS analysis to study benchmark points in our simplified models producing cross section limits, limits in the mediator-dark matter mass plane, 
and cross section versus mass limits. 

The introduction of the simplified models greatly increases the number of free parameters which enter searches at the LHC. Previous iterations of
experimental results focused on the regime in which the mediator is assumed to be heavy, such that an effective field theory description is valid. 
In this setup one has to constrain one parameter per operator, the Wilson coefficient $C_i$. However at the LHC energies many interesting scenarios 
occur in which the mediator can be produced on-shell, which makes the introduction of simplified models a useful tool. Future iterations of LHC searches 
have the harder task of presenting results in this five dimensional plane.  

If the simplified model is extended to include coloured degrees of freedom, modifications can occur in the production of the mediating particle. This is particularly 
relevant for the scalar and pseudo scalar cases, which proceed at the one-loop level. Heavy coloured physics can couple directly to the mediator, and result in an additional 
5 dimensional contribution to the Lagrangian. We investigated the impact of this term for the scalar mediated case and found that if the mediator-top coupling is damped then 
significant contributions to the production cross section can arise from these terms. This is analogous to the situation in the SM, in which the light bottom quark loops are much smaller than the top quark loops (which can be treated in an effective field theory approach). 

We find that limitations of direct and indirect detection experiments, i.e. velocity suppression and loop-suppressed couplings to Standard Model particles, can be overcome by LHC searches. Thus, the LHC provides a complementary coverage of the dark sector parameter space with respect to low-energy experiments. Importantly, if the invisible particle is not stable on cosmological time scales the LHC can be the only experiment to probe the dark sector.

The search for dark matter whether in direct, indirect, or collider experiments represents one of the most fascinating, and challenging 
goals for physics this century. The LHC is about to enter a new era with the start of Run II, and with this the evolution of the LHC searches 
to incorporate more complete UV models and include the region in which EFT assumptions breakdown, is a natural progression. The simplified 
models we have discussed here provide a pathway to achieving this. 
\medskip

\noindent {\bf Note added}:
When this paper was being finalised, Ref.~\cite{Buckley:2014fba}  was posted on the ArXiv, which also considers scalar and pseudo-scalar 
mediators in the gluon fusion channel and treats $\Gamma_{\rm MED}$ as a free parameter.

\section*{Acknowledgments}
\noindent We would like to thank Brian Batell, Oliver Buchmueller, Albert De Roeck, Patrick Fox, Christopher McCabe, Tim Tait and Christopher Wallace for valuable discussions.
The research of VVK and MS is supported by STFC through the IPPP grant and for VVK by the Wolfson Foundation and Royal Society.

\appendix
\label{app:amplitudes}

\section{Amplitudes for $gg\rightarrow S+g$ and $q\overline{q}\rightarrow S+g$ }
In this section we present the helicity amplitudes for $gg\rightarrow S+g$ and $q\overline{q}\rightarrow S+g$ in both the full and effective 
field theories. 
Helicity amplitudes are defined in terms of $u_{\pm}(k_i)$ where $u$ represents a Weyl spinor of momentum $k_i$, with 
either positive or negative helicity. Basic spinor products are then defined as,
\begin{eqnarray}
\spa i.j = \langle i^- | j^+\rangle= \overline{u}_-(k_i)u_+(k_j), \\
\spb i.j =  \langle i^+ | j^-\rangle = \overline{u}_+(k_i)u_-(k_j), 
\end{eqnarray}
Kinematic invariants are constructed from products of the above quantities, 
\begin{eqnarray}
\spa i.j\spb j.i = 2 k_i k_j = s_{ij}
\end{eqnarray}
Spinor strings are defined as follows, 
\begin{eqnarray}
\spab i.P_{kl}.j = \spab i.(k+l).j = \spa i.k \spb k.l+\spa i.l\spb l.j
\end{eqnarray}

We begin by presenting the $gg\rightarrow S+g$ amplitude with a propagating fermion of mass $m_f$ in the loop. We decompose the amplitude in terms of a kinematic primitive amplitude, and normalization factors as follows,
\begin{eqnarray}
A_4(S,1_g^{h_1},2_g^{h_2},3_g^{h_3})=N_c(N_c^2-1)\frac{g_f m_f^2 g_w}{16\pi^2 M_W }\left(\frac{g_s}{\sqrt{2}}\right)^3\mathcal{A}(S,1_g^{h_1},2_g^{h_2},3_g^{h_3})
\end{eqnarray}
Of the possible helicity orderings two can be chosen ($+++$, $-++$) from which all remaining amplitudes can be obtained from conjugation and bose symmetries. 
The $+++$ amplitude is given in terms of the following box ($D_i$) and triangle $(C_i)$ scalar integrals 
\begin{eqnarray}
\mathcal{A}(S,1_g^{+},2_g^{+},3_g^{+})&=&\bigg(\frac{\spb2.1\spb3.1(4m_f^2-s_{123})}{\spa 2.3}D_1(s_{13},s_{12},m_f^2) +
\frac{s_{12}+s_{13}}{\spa1.2\spa1.3\spa2.3}(4m_f^2-s_{123})C_1(s_{23},s_{123},m_f^2)   \nonumber\\&&+\{ 1 \leftrightarrow 3\}   +\{ 1 \leftrightarrow 2\} \bigg)+ \frac{s_{123}}{\spa1.2\spa1.3\spa2.3}
\end{eqnarray}
Here $D_1(s,t,m_f^2)$ represents a box integral with a single off-shell leg ($s_{123}$), which is specified completely by $s$ and $t$ channel invariants, $C_1(s,t,m_f^2)$ represents the triangle integral with two legs $s$ and $t$ off-shell. 
The second helicity amplitude required also contains bubble integrals $B_i$, 
\begin{eqnarray}
\mathcal{A}(S,1_g^{+},2_g^{+},3_g^{+})&=&-\frac{\spa 1.2\spa1.3}{\spa 2.3^3}(4 s_{12}s_{13}+12m_f^2-s_{23}^2)D_1(s_{13},s_{12},m_f^2)
+2\frac{(s_{12}+s_{13})\spb3.2(4m_f^2-s_{23})}{\spa2.3^2\spb2.1\spb3.1}C_1(s_{23},s_{123},m_f^2)
\nonumber\\&&+\bigg\{\frac{\spa1.2\spb2.3^2(4m_f^2-s_{23})}{\spa2.3\spb3.1}D_1(s_{12},s_{23},m_f^2)+ \frac{4\spa1.2\spa1.3 s_{13}}{\spa2.3} C_2(s_{13},m_f^2)   \nonumber\\&&
+\bigg(-4\frac{\spa1.2\spa1.3\spab3.P_{12}.3}{\spa2.3^3} +2\frac{s_{13}\spa1.3\spb3.2^2}{\spa2.3\spb2.1\spab3.P_{12}.3} -2\frac{s_{23}\spb2.3^3}{\spb2.1\spb3.1\spab3.P_{12}.3}\nonumber\\&&-4m_f^2\bigg(2\frac{s_{13}\spa1.3\spb3.2}{\spa2.3^2\spb2.1\spab3.P_{12}.3}+4\frac{\spa1.3\spb3.2^3}{\spa2.3\spb2.1\spab3.P_{12}.3}+2\frac{\spb3.2^3}{\spab3.P_{12}.3\spb2.1\spb3.1}\bigg)\bigg)C_1(s_{12},s_{123},m_f^2) \nonumber\\&& +8\frac{\spa1.2\spa1.3\spb3.2(s_{13}+2 s_{123})}{\spa2.3^2\spab3.P_{12}.3^2}\bigg(B_1(s_{12},m_f^2)-B_1(s_{123},m_f^2)\bigg)+ \{ 2 \leftrightarrow 3\} \bigg\} \nonumber\\ &&
-\frac{\spb3.2^2(s_{12} s_{13}+s_{23}^2)}{\spa2.3\spb2.1\spb3.1\spab3.P_{12}.3\spab2.P_{13}.2}
\end{eqnarray}
Here we have introduced additional triangle integrals $C_2(s,m_f^2)$ which represent topologies with one off-shell leg ($s$), and 
bubble integrals $B_1(s,m_f^2)$  which depend on a scale $s$.
The basis integrals in the above equations can be easily evaluated using public packages, our implementation in MCFM uses QCDLoop~\cite{Ellis:2007qk}. 

The amplitude for  $q\overline{q}\rightarrow S+g$ can be written as follows, 
\begin{eqnarray}
A_4(S,1_{\overline{q}}^{h_1},2_q^{-h_1},3_g^{h_3})=(N_c^2-1)\frac{g_f m_f^2 g_wg_s^3}{16\pi^2 M_W }\mathcal{A}(S,1_{\overline{q}}^{h_1},2_q^{-h_1},3_g^{h_3})
\end{eqnarray}
here the primitive amplitude can be defined in terms of one helicity configuration ($-++$) with all other configurations obtainable from line reversal and conjugation symmetries. Our primitive amplitude is 
\begin{eqnarray}
\mathcal{A}(1_{\overline{q}}^-,2_q^+,3+)&=&2\frac{\spb2.3^2}{\spb2.1}(\spab3.P_{12}.3-4m_f^2)C_1(s_{12},s_{123},m_f^2)\nonumber\\&&-4\frac{\spa1.2\spb3.2^2}{\spab3.P_{12}.3^2}(B_1(s_{12},m_f^2)-B_1(s_{123},m_f^2)
+4\frac{\spb3.2^2}{\spb2.1\spab3.P_{12}.3}
\end{eqnarray}

In addition to the amplitudes presented above, we will also need the $m_f\rightarrow \infty$ limit, which corresponds to amplitudes
computed in the effective field theory. These amplitudes have been computed in ref.~\cite{Badger:2004ty}, and have the following form,
\begin{eqnarray}
\mathcal{A}_{m_f\rightarrow \infty}(S,1_g^+,2_g^+,3_g^+)=\frac{s_{123}^2}{\spa1.2\spa2.3\spa 3.1} \\
\mathcal{A}_{m_f\rightarrow \infty}(S,1_g^+,2_g^-,3_g^-)=\frac{\spa2.3^2}{\spa1.2\spa 3.1} 
\end{eqnarray}

\bibliography{ref}
\bibliographystyle{ArXiv}

\end{document}